\documentclass[preprinttwo, twocolumn]{aastex63}
\usepackage{amsmath,amssymb,lmodern,mathrsfs,ulem,mathtools,bm}
\usepackage{graphicx}
\usepackage{enumitem}
\usepackage{natbib}
\usepackage{xcolor}
\usepackage{makecell}
\usepackage{hyperref}
\usepackage{lipsum}
\usepackage[modulo]{lineno}
\usepackage[caption=false]{subfig}

\newcommand{\divrm}{\mathrm{div}}

\definecolor{forest}{RGB}{34,139,34}

\newcommand{\rev}[1]{{#1}}

\journalinfo{Submitted to ApJ}

\begin{document}

\received{March 18, 2024}
\revised{August 8, 2024}
\accepted{August 23, 2021}

\title{Simulation of Thermal Nonequilibrium Cycles in the Solar Wind}

\author[0000-0001-8517-4920]{Roger B. Scott}
\affiliation{\rm Space Science Division, US Naval Research Laboratory, Washington, DC 20375, USA}

\author[0000-0003-4739-1152]{Jeffrey W. Reep}
\affiliation{\rm Institute for Astronomy, University of Hawai'i at M\=anoa, Pukalani, HI 96768}

\author[0000-0002-4459-7510]{Mark G. Linton}
\affiliation{\rm Space Science Division, US Naval Research Laboratory, Washington, DC 20375, USA}

\author[0000-0002-3300-6041]{Stephen J. Bradshaw}
\affiliation{\rm Department of Physics and Astronomy, Rice University, Houston, TX 77005, USA}

\begin{abstract}
\noindent Thermal nonequilibrium (TNE) is a condition of the plasma in the solar corona in which the local rate of energy loss due to radiation increases to the point that it cannot be sustained by the various heating terms acting on the plasma, precluding the existence of a steady state.
The limit cycles of precipitation and evaporation that result from TNE have been simulated in 1D models of coronal loops, as well as 2D and 3D models of the solar chromosphere and lower corona.
However, a careful study of TNE in the solar wind has not been performed until now.
Here we demonstrate that for suitable combinations of local and global heating rates it is possible for the plasma to exhibit a TNE condition, even in the context of a transonic solar wind with appreciable mass and energy fluxes.
This implies limits on the amount of foot-point heating that can be withstood under steady-state conditions in the solar wind, and may help to explain the variability of solar wind streams that emanate from regions of highly concentrated magnetic flux on the solar surface.
The implications of this finding pertain to various sources of high-density solar wind, including plumes that form above regions of mixed magnetic polarity in polar coronal holes and the slow solar wind (SSW) that emanates from coronal hole boundaries.
\\
\end{abstract}

\section{Introduction}\label{Section::Introduction}

The condensation and precipitation of coronal plasma, commonly referred to as ``coronal rain,'' has been a ubiquitous feature in H-$\alpha$ observations of the solar corona ever since the first birefringent filter telescope was designed and constructed by \citet{Evans:1949, Evans:1958} at the High Altitude Obseratory in Boulder, CO.
Initially this phenomenon was discussed in the context of coronal prominences \citep[see, e.g., ][]{Newkirk:1957, Kawaguchi:1970, Leroy:1972}, but was later found in observations of coronal loops by such authors as \citet{Schrijver:2001}, using the TRACE instrument \citep{Handy:1999}, and \citet{Degroof:2004}, using SOHO/EIT \citep{Delaboudiniere:1995} and, later, the Big Bear Solar Observatory \citep{DeGroof:2005}.
In the era of AIA/SDO \citep{Lemen:2012} such observations have become common, particularly in coronal active regions where the stored magnetic energy is large. 
Recently, \citet{Mason:2022} performed a survey of 241 solar flares from 2011 to 2018 and determined that most ($>90\%$) X-class flares and a majority ($>70\%$) of M-class flares exhibited coronal rain along the post-flare magnetic arcade over a period of several hours following the peak in flare brightness.
This suggests that the conditions under which coronal condensates form are ubiquitous in the solar corona, especially where there is strong heating, as is common in solar active regions \citep{Fisher:1998}.

Additionally, there is evidence that coronal rain may occur preferentially near magnetic null points and along field lines that pass near to them.
\citet{Mason:2019} reported on observations of coronal rain along the separatrix surfaces of coronal null points above embedded magnetic bipoles observed by AIA/SDO.
They even observed condensations forming above the (observationally inferred) location of the null point, along its open spine.
This suggests that such condensates may not develop exclusively in the closed corona, as has typically been assumed.
Some authors have even gone as far as to link the formation of coronal rain with the process of interchange reconnection \citep{Li:2020}, although the causal mechanism underpinning this connection remains to be determined.

The first computational model of coronal condensation was developed by \citet{Antiochos:1991} to describe the physics of prominence formation.
Building on previous results by \citet{Serio:1981}, they showed that for sufficiently strong foot-point heating rates the density at the apex of a coronal loop can increase to the point that the radiated power exceeds the power supplied by all other sources of heating, after which the plasma undergoes a localized thermal runaway that ultimately leads to catastrophic cooling and condensation.
This phenomenon, which is now referred to as thermal nonequilibrium, or TNE, has since been confirmed in field aligned (1D) simulations by \citet{Antiochos:1999}, \citet{Antolin:2008, Antolin:2010}, \citet{Downs:2016}, \citet{Froment:2018}, \citet{Johnston:2019}, and others.
Moreover, \citet{Klimchuk:2019o} have developed scaling laws for determining the availability of steady state solutions on closed coronal loops with prescribed heating, and these have proven to be highly predictive of TNE conditions in numerical simulations.

Although the majority of TNE studies have been performed in 1D models of closed coronal loops, the condensation of coronal plasma has also been observed in 2.5D and 3D simulations by \citet{Fang:2015}, \citet{Kohutova:2020}, \citet{Li:2022}, \citet{Antolin:2022}, and others.
In a few cases, these have been observed to form along ``open'' field lines that extend from the model chromosphere to the top of the numerical domain.
This suggests that TNE is insensitive to magnetic connectivity, meaning that it may occur in coronal holes or near coronal hole boundaries.
However, because the spatial domains in these simulations typically don't extend beyond about $50\,\rm Mm$ above the solar surface, they do not support an outward-flowing solar wind with a sonic point.
As a result, the gas dynamics along open field lines in such models may not be representative of the conditions found in the regions whence the solar wind emanates.

One exception is the study by \citet{Schlenker:2021a}, who performed a 2.5D simulation of TNE conditions in a magnetic helmet streamer with a self-consistent, transonic solar wind in the adjacent coronal hole region.
They found that certain values of the foot-point heating rate caused coronal condensates to form along the flank of the helmet streamer, and that the critical condition predicted by \citet{Klimchuk:2019o} was approximately recovered when the loop length was replaced with the gravitational scale height to establish the natural length scale of the system.
Notably, some of the condensates observed by \citet{Schlenker:2021a} appeared to be raining down along open field lines that were outside of the helmet streamer separatix surface (i.e., the open-closed boundary); however it was not clear from their simulations whether the condensates originated in the open field region or were released into it following magnetic reconnection at the helmet streamer apex.

In this paper we report on the formation of coronal rain and the emergence of TNE-driven limit cycles of precipitation and evaporation (hereafter ``TNE cycles'') in numerical simulations of the solar wind.
Our model solves the time-dependent field-aligned Navier-Stokes equations from the chromosphere, through the transition region and lower corona, and into the extended corona and heliosphere, with empirical heating, radiative cooling, and thermal conduction.
As this model explicitly assumes an open magnetic geometry, which naturally supports a transonic outflow, the condensates that form within the model, which are found in the subsonic region between the solar surface and the sonic point, are unambiguously embedded in and consistent with the conditions of the solar wind.

In the following section (\ref{Section::Model}) we briefly describe the design of our numerical model.
Then, in sections \ref{Section::Equilibrium} and \ref{Section::Onset}, we describe the equilibrium conditions and subsequent loss of equilibrium as a consequence of enhanced foot-point heating.
The TNE cycles that emerge following the loss of equilibrium are discussed in sections \ref{Section::Phases} and \ref{Section::Periodicity}. 
In section \ref{Section::Signatures} we discuss the implications of TNE cycles to heliospheric observations, before concluding with a summary discussion in section \ref{Section::Discussion}.

\section{Model Design}\label{Section::Model} 

We use the HYDrodynamics and RADiation code \citep[HYDRAD,][]{Bradshaw:2013} to solve the field-aligned Navier-Stokes equations in 1D for the mass density of ions and electrons $\rho_{i,e}$, bulk (ion) velocity $u$, and ion and electron pressures $p_{i,e}$, along a radially expanding magnetic flux tube with open (extrapolated) boundary conditions at the heliospheric end of the domain.
The ions comprise a single species representing a weighted average of singly-ionized hydrogen and helium, with a mean ion mass of $m_i = 2.17\times10^{-24}\,\rm g$, while the electron mass is $m_e = 9.11\times10^{-28}\,\rm g$, and we assume charge neutrality, with equal ion and electron particle densities ($n = n_i = n_e$), where the particle densities are related to the mass densities as $n_{i,e} = \rho_{i,e} / m_{i,e}$.
We define the temperatures of the two species in terms of their partial pressures through the ideal gas law $p_{i,e} = n k_B T_{i,e}$, where $k_B = 1.38\times10^{-16}\,\rm erg\,K^{-1}$ is Boltzmann's gas constant. 

The numerical domain is described by the spatial coordinate $s$ that spans a height of $0 \le s \le 30 \, R_\odot$ above the solar surface, with the radius of the sun taken to be $R_\odot = 700 \rm \, Mm$.
The adaptively-refined grid comprises a minimum of 512 cells that are equally spaced in $\ln(r/R_\odot)$, where $r(s) = s+R_\odot$ is the heliocentric radius associated with a given position $s$.
During refinement, each cell can be subdivided into a pair of equally-sized sub-cells, up to a maximum of 16 times, for a minimum grid size of $\Delta s_{\rm min}\simeq0.1\,\rm km$ at $s=0$.
Similarly, during de-refinement cells are merged pair-wise, so that the largest fractional change in value between adjacent cells for any primitive variable is kept in the range of $5\% - 10\%$.

The modeled system of equations and boundary conditions are identical to the description in \citet{Scott:2022a} except that a single-fluid condition has been imposed by increasing the collisional coupling, so that the ions and electrons have equal temperatures ($T=T_i=T_e$) and, therefore, equal pressures ($p = p_i = p_e$).
As seen in the simulations performed by \citet{Scott:2022a}, which allow for separate thermal evolution of the two species, $T_i \approx T_e$ up to about one pressure scale height ($h_{p} \simeq 90\, \rm Mm$) above the solar surface, which is higher than the formation height of the condensates in this study, so this choice has little impact on the dynamics presented here.

Given the single-temperature condition, it is convenient to restate the combined energy equations of the ions and electrons in terms of the total pressure $P = p_i + p_e = 2 p$, which evolves according to 
\begin{equation}\label{Equation::energy}
    \partial_t P + {\rm div}\left(P u\right) = \left (\gamma - 1 \right ) \left ( \mathcal{H} + \mathcal{R} + \dot{Q}_c - P\, {\rm div}(u) \right ),
\end{equation}
where ${\rm div}(\circ):= (1/A)\, \partial_s (A\,\circ)$ is the 1D divergence operator along the expanding flux tube, $\partial_s$ is the directional derivative along $s$, and $A(s) = A_0 r(s)^2/R_\odot^2$ is the parameterized cross-sectional area of a radial flux tube in a spherically-symmetric magnetic field.
The four terms on the right-hand side (RHS) of this expression are the imposed external heating $\mathcal{H}$, radiative energy loss $\mathcal{R}$, and the heating and cooling due to thermal conduction and adiabatic compression and expansion.

The conductive heating and cooling, $\dot{Q}_c = -\, {\rm div}\, (f_c)$, depends on the heat flux $f_c$, which comprises two terms: the collisional heat flux $f_\kappa$ given by Fourier's law, and the free-streaming limit of the electron energy flux $f_{v_e}$.
These are defined as 
\begin{equation}
    f_\kappa = -\kappa(T) \partial_s T \quad \text{and} \quad f_{v_e} = \frac{1}{4} m_e n v_e^3,
\end{equation}
where $v_e = \sqrt{k_B T / m_e}$ is the thermal speed of the electrons and $\kappa(T) = \kappa_0 T^{5/2}$ is the Spitzer-H\"arm conductivity \citep{Spitzer:1953}. 
The single-fluid coefficient of thermal conduction $\kappa_0$ (which is the sum of ion and electron conductivities) is set implicitly by its value at $T=10^6\,\rm K$, which we call $\kappa_6 \equiv \kappa( {\scriptstyle T = 10^6\,\rm K} ) = 8.12\times10^{8}\rm\,erg\, cm^{-1} K^{-1} s^{-1}$.
The stream-limited heat flux is then 
\begin{equation}
    f_c = \frac{f_\kappa f_{v_e}}{\sqrt{ f_\kappa^2 + f_{v_e}^2 }}.
\end{equation}
This construction guarantees that $f_c$ is parallel to $f_\kappa$, and tends to $f_\kappa$ in the limit that $f_{v_e}$ is small, while its value is bounded by $f_{v_e}$ \citep{Cowie:1977}.

The radiative energy loss is modeled as an empirical function of temperature and density, which is prescribed to fall smoothly to zero below a threshold temperature of $T_{0} = 2\times10^4\, \rm K$, emulating the behavior of an optically thick plasma at the interface between the transition region and the model chromosphere.
The radiative loss function is given by $\mathcal{R}(n,T) = -n^2 \Lambda(T)$ where $\Lambda(T)$ is the coronal emissivity, which is modeled as a piece-wise continuous function of $T$ as given in \citet{Klimchuk:2008}.
\rev{The specific form of $\Lambda(T)$ is a collection of power laws defined over specific temperature intervals that span the whole range of chromospheric and coronal temperatures, with spline points at $\log_{10}T = \{4.97, 5.67, 6.18, 6.55, 6.90, 7.63\}$. 
For the purposes of this study we are concerned with the emissivity below and up to about $10^6\,\rm K$, for which there are three relevant intervals.
In particular, for temperatures below $10^{4.97}\,\rm K$ the emissivity increases at $T^2$, while for}
temperatures in the range $4.97 < \log_{10} T < 6.18$ the emissivity is 
\begin{equation} 
    \Lambda(T) = \Lambda^\prime \times (T/T^\prime)^b
\end{equation}
where $T^\prime = 10^{5.67}\,\rm K\simeq 0.47\,\rm MK$, with
\begin{equation}
    b = \begin{cases}
-1             \quad  \text{for} & T < T^\prime \\
\hspace{7pt} 0 \quad  \text{for} & T > T^\prime,
\end{cases} 
\end{equation}
and $\Lambda^\prime = 1.9\times10^{-22}\, \rm erg \, cm^{3}\, s^{-1}$.
At the critical temperature $T^\prime$ the form of the emissivity changes from a constant value to an inverse function of temperature, so that for a given value of $n$ the radiative energy loss increases with decreasing temperature below $T^\prime$.

We prescribe the form of the coronal heating to be a superposition of two decaying exponential functions, with length scales of $\ell_g = R_\odot$ and $\ell_b = 0.01\, R_\odot$, so that
\begin{equation}\label{Equation::Heating}
    \mathcal{H}(s) = A(s)^{-1} \times \left ( \mathcal{H}_b e^{-(s-s_0)/\ell_b} + \mathcal{H}_g e^{-(s-s_0)/\ell_g} \right ).
\end{equation}
The prefactor of $1/A$ is a normalization that is included to ensure that the energy deposited per unit length along the flux tube is independent of its cross-sectional area, while $s_0 \equiv 10\,\rm Mm$ is the reference height for the base of the transition region (TR) at the start of the simulation.
Below $s_0$, the model chromosphere extends down to the base of the numerical domain and provides a large reservoir of material below the TR, which insulates the dynamics within the TR from any effects due to the closed boundary conditions at $s=0$.

The two exponential terms in Eq. \eqref{Equation::Heating} represent the large-scale (global) heating rate, which primarily dictates the kinetic energy flux of the solar wind beyond the sonic point, and the small-scale (foot-point) heating rate, which primarily dictates the temperature and density of the plasma in the transition region and lower corona. 
With the chosen normalization the volume-integrated heating also separates easily into the two contributions,
\begin{equation}
    \int_C \mathcal{H}(s) dV = \int_{s_0}^\infty \mathcal{H}(s) A(s) ds = \mathcal{H}_b \ell_b + \mathcal{H}_g \ell_g.
\end{equation}
The amplitude of the global heating rate is fixed at $\mathcal{H}_g = 2\times10^{-6}\, \rm erg\, cm^{-3}\, s^{-1}$, while the foot-point heating is set to $\mathcal{H}_b = \alpha \times \mathcal{H}_g \ell_g / \ell_b$.
The heating ratio $\alpha$ is treated as a free parameter, which is used to control the ratio of foot-point to global heating, with $\alpha=1$ corresponding to the case $\mathcal{H}_b\ell_b = \mathcal{H}_g \ell_g$.
Because the two heating terms are additive, the minimum value of the heating rate at the base of the corona, when $\alpha=0$, is $\mathcal{H}_g$, while values of $\alpha \gtrsim 1$ correspond to foot-point heating rates in excess of $100 \times \mathcal{H}_g$, for which the contribution from the global heating is negligible at heights below $s\sim s_0 + \ell_b$.

The simulation is initialized with a hydrostatic atmosphere and no foot-point heating ($\alpha=0$) and allowed to relax until a steady-state wind solution has formed, as discussed in \citet{Scott:2022a}.
We then set $\alpha=1$ and let the simulation relax until a new steady-state wind solution has formed.
From there we increase $\alpha$ to incrementally larger values, always allowing sufficient time for the solution to settle to a steady-state if one is available.
In total we impose 12 different values of $\alpha$ in the range $0\leq \alpha \leq 10$, and the resulting dynamics are discussed in the following sections.

\section{Equilibrium Conditions}\label{Section::Equilibrium}

We will ultimately be concerned with the loss of equilibrium and subsequent thermal runaway that occur with increased rates of foot-point heating. 
To understand the conditions under which this occurs, we will first consider the structure of steady-state solutions subject to marginal rates of foot-point heating ($\alpha \lesssim 1$), after which we can discuss the availability of such solutions as the foot-point heating rate is increased.

In the absence of additional foot-point heating, when $\alpha=0$, the base of the TR, where $\mathcal{R} \rightarrow 0$ as $T\rightarrow T_{0}$, is located at $s \simeq s_0$.
Above this height the temperature increases rapidly within the TR to several $10^5\,\rm K$ over a distance of $<1\,\rm Mm$, while the electron number density decreases by roughly an order of magnitude over the same distance.
Conversely, the pressure does not change significantly across the TR; however, the magnitude of the pressure gradient decreases dramatically in response to the decrease in density, which dictates a proportionate reduction in the gravitational force.
Above the TR, the temperature then rises gradually with height in the corona to a maximum of $\sim1.5\,\rm MK$ at a height of $s\sim1.5 R_\odot$ while the density and pressure fall off monotonically with height in the corona.

The location of the temperature peak is dictated by the external heating and radiative loss rates, both of which decrease monotonically with height -- the heating rate decaying exponentially while the radiative cooling rate scales with the square of the density. 
But because $n$ decreases faster than $\mathcal{H}$ in the lower corona and then ultimately falls off as $s^{-2}$ in the heliosphere, the external heating rate per particle $\mathcal{H} / 2n$ peaks at a height of about $1R_\odot$ above the solar surface, corresponding to the global heating scale height $\ell_g$. 
At (and beyond) this height the radiative cooling rate is negligible, so the energy budget in the region $s\gtrsim R_\odot$ is effectively dominated by thermal conduction, external heating, and, to a lesser degree, adiabatic expansion.

To understand why $\mathcal{H}/2n$ and $T$ peak at similar heights in the corona, it is convenient to restate the energy equation \eqref{Equation::energy} in terms of the thermal energy per particle $\epsilon = k_B T / (\gamma - 1)$.
To this end we first expand the LHS of Eq \eqref{Equation::energy} as
\begin{equation}
    \divrm (Pu) = \rho u \partial_s (P/\rho) + (P/\rho)\, \divrm (\rho u).
\end{equation}
Then, since $\divrm(\rho u) = \partial_t \rho = 0$ under steady-state conditions, while the ratio $P/\rho = 2 k_B T / m_i$, it follows that
\begin{equation}
    \frac{1}{2n} \divrm(Pu) = u \partial_s (k_B T).
\end{equation}
Consequently, the steady-state governing equation for $k_B T$ (and, hence, $\epsilon$) is
\begin{equation} \label{Equation::temperature}
   \frac{u}{\gamma-1} \partial_s (k_B T) = \left ( \mathcal{H} + \mathcal{R} + \dot{Q}_c \right )/2n - k_B T \, {\rm div}(u).
\end{equation}
This expression is equivalent to equation \eqref{Equation::energy}, but demonstrates an important property of the temperature, which distinguishes it from volumetric quantities like pressure and density:
namely, while $P$ evolves in response to a velocity-dependent flux, $T$ is transported by the advective derivative ($u\partial_s)$, meaning that $\rm div(u)$ only enters through the adiabatic heating term on the RHS of Eq. \eqref{Equation::temperature}.
Consequently, where the various terms on the RHS of Eqs. \eqref{Equation::energy} and \eqref{Equation::temperature} are small, $u\partial_s T$ must also be small, and where $\partial_s T = 0$, the sum of all local heating and cooling must also be zero.

While the governing equation \eqref{Equation::temperature} for $T$ allows that $\partial_s T \rightarrow 0$ where the net rates of heating and cooling are zero, it does not directly imply anything about the local variation in $\mathcal{H}$ at that location. 
Indeed, since $\dot{Q}_c$ is itself independent of $n$ in the Spitzer-H\"arm formulation, there is no reason to expect that heating and cooling terms should all add to zero at the point where $\mathcal{H}/2n$ is maximal. 
Rather, that result is a consequence of the stream-limited form of the conductive heat flux, which manifests as a cooling term that scales as $\dot{Q}_c/2n \sim \left ( T^{3/2} \times \partial_s \ln n \right )$ at heights $\lesssim \ell_g$.
And because $f_{v_e} \propto n m_e {v_e}^3$ decreases rapidly with height, the conductive heat flux in the extended corona is always stream limited except where $\partial_s T$ is very small.
So as adiabatic cooling overtakes external heating at larger heights in the corona the temperature there must decrease with height in order that $f_c > 0$, as needed to maintain the internal energy of the expanding plasma.
It follows, therefore, that $T$ must increase with $\mathcal{H}/2n$ in the lower corona while also decreasing where $\mathcal{H}/2n$ is small, meaning that their peaks occur at similar heights.

\begin{figure}
\centering
\includegraphics[width=\linewidth]{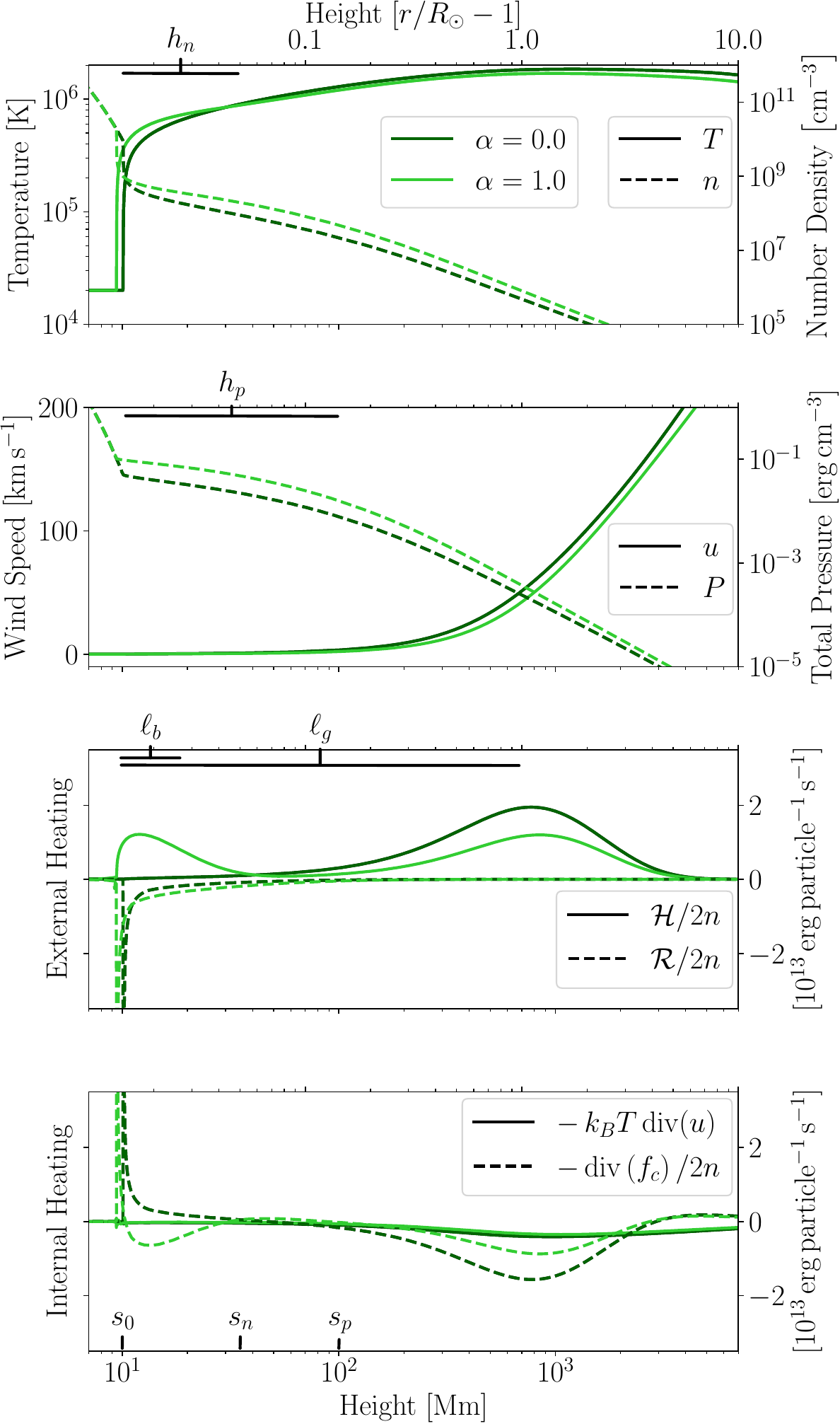}
\caption{Profiles of plasma number density and temperature (top most panel), pressure and wind speed (second panel), and rates of heating per particle (lower two panels) are depicted for two steady state solutions with different foot-point heating rates. The values of $\alpha \in \{0.0, 1.0\}$ correspond to heating rates of $\mathcal{H}_b \in \{0.0, 2.0\} \times 10^{-4}\, \rm erg\, cm^{-3}\, s^{-1}$. 
The global heating in each case is $\mathcal{H}_g = 2.0\times10^{-6}\, \rm erg\, cm^{-3}\, s^{-1}$. }
\label{Figure::Baseline}
\end{figure}

The various terms in Equation \eqref{Equation::temperature} are depicted by the black curves in Figure \ref{Figure::Baseline} for a typical solar wind solution without foot-point heating ($\alpha=0$), and for a slower, denser solar wind solution with moderate foot-point heating ($\alpha=1$).
The temperature, density, wind speed, and total pressure are shown in the two upper panels.
The heating (cooling) terms are shown in the lower two panels, and these are separated into external heating rates (prescribed heating and radiative cooling) and internal heating rates (adiabatic expansion and thermal conduction).

Above the temperature peak $f_c$ is always directed away from the sun, as is $P u$, so any heat deposited at or above that height is necessarily transported outward, either by conduction or advection, or lost to radiation.
Below the temperature peak, the heat flux is directed inward toward the sun, and serves to transport heat energy down from the upper corona to support the radiative energy losses that dominate the energy budget in the TR.
And because $P\,{\rm div}(u)$ is small for $s_0 < s < \ell_g$, the primary energy balance there is dictated by $\mathcal{H}+\mathcal{R} + \dot{Q}_c \simeq 0$. 

When $\alpha = 0$ there is an extended region of excess cooling ($\mathcal{H} + \mathcal{R} < 0$) that spans from $s_0$ to $s_{n} \equiv s_0 + h_{n}$, where $h_{n}\simeq25\,\rm Mm$ is the density scale height at the base of the corona.
Consequently, $\dot{Q}_c$ acts as a heating term in the region $s_0 < s \lesssim s_{n}$, and then changes to a cooling term in the $\mathcal{H}$-dominated region $s_{n} \lesssim s \lesssim \ell_g$.
However, when $\alpha = 1$ the additional external heating at the base of the corona ($s \lesssim \ell_b$) creates an additional region of excess heating between the TR and $s_{n}$, with the result that $\dot{Q}_c$ takes on additional complexity, becoming a cooling term in the region $s_0 < s \lesssim \ell_b$, before reverting to a heating term in the weakly $\mathcal{R}$-dominated region between $s_{n} \lesssim s \lesssim s_{p}$.
Here $s_{p} \equiv s_0 + h_{p}$, and $h_{p} \simeq 90\,\rm Mm$ is the pressure scale height at the base of the corona. 
Beyond $s_{p}$, $\dot{Q}_c$ then returns to a cooling regime in the $\mathcal{H}$-dominated region $s_{p} \lesssim s \lesssim \ell_g$.

While the region of conductive cooling in the low corona dissipates some of the energy from the additional foot-point heating, the structure of the TR is such that all of the heat deposited within it must be radiated away locally, so any increase in heating there necessitates an increase in $\mathcal{R}$, and hence an increase in $n$ (since the emissivity at the base of the TR is fixed). 
Consequently, in the case of $\alpha=1$ the base of the TR is shifted deeper into the chromosphere compared to the case of $\alpha=0$, which increases the density at the base of the TR and at all heights above it, through to the extended corona and heliosphere. 
But since the energy flux of the solar wind -- which is dominated by $\rho u^3$ -- is effectively independent of the foot-point heating, being dictated instead by the global heating rate, the increase in density ($\Delta n$) that results from the application of additional foot-point heating has the additional effect of decreasing the wind speed by roughly a factor of $\Delta u \propto (\Delta n)^{-1/3}$.
Similarly, the mass flux of the solar wind is increased by a factor of $\Delta \rho u \propto (\Delta n)^{2/3}$, while the temperature of the asymptotic wind is decreased by a factor of the global heating rate per unit mass flux, meaning that $\Delta T \propto (\Delta n)^{-2/3}$.
This behavior is consistent with the previous finding by \citet{Grappin:2011} that increased foot-point heating results in a slower, cooler, and denser solar wind.

\begin{figure}
\centering
\includegraphics[width=\linewidth]{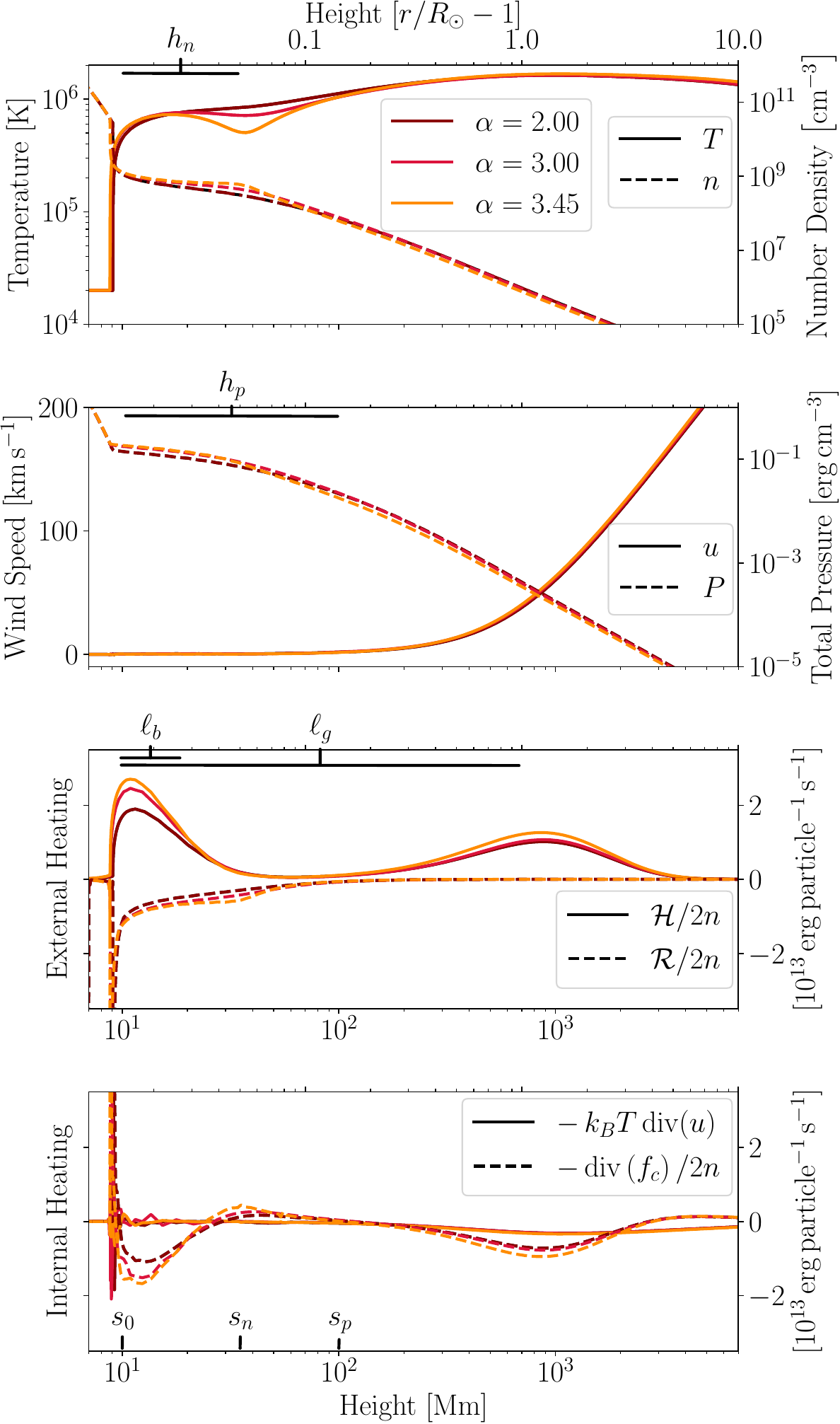}
\caption{Additional profiles of plasma number density and temperature (top most panel), pressure and wind speed (second panel), and rates of heating per particle (lower two panels) for three steady state solutions with increased foot-point heating rates. The global heating in each case is unchanged from $\mathcal{H}_g = 2\times10^{-6}\, \rm erg\, cm^{-3}\, s^{-1}$. The values of $\alpha \in \{2.0, 3.0, 3.45\}$ correspond to foot-point heating rates of $\mathcal{H}_b \in \{4.0, 6.0, 6.9\} \times 10^{-4}\, \rm erg\, cm^{-3}\, s^{-1}$. }
\label{Figure::Marginal}
\end{figure}

The effects of further increases in the foot-point heating are shown in the black, orange, and red curves in Figure \ref{Figure::Marginal} for the cases of $\alpha = \{2.0, 3.0, 3.45\}$.
Looking at the heating and cooling rates in the bottom two panels there is little qualitative change in the profiles of $\mathcal{H}/2n$ at heights $s\gtrsim\ell_b$ as $\mathcal{H}_b$ is increased, although there is a visible enhancement in the radiative cooling term $\mathcal{R}/2n$ near $s_{n}$.
This coincides with an increase in the density, which falls off more slowly with height between the TR and $s_{n}$ as $\mathcal{H}_b$ is increased.
The conductive cooling at the base of the corona also appears to increase with increasing values of $\alpha$ but the effect becomes smaller with each subsequent increase as $\mathcal{H}_b/2n$ increases more slowly than $\mathcal{H}_b$, owing to the corresponding increases in $n$.
Meanwhile, the conductive heating near $s_{n}$ appears to increase in response to each increase in $\mathcal{R}/2n$ at that height.

The most obvious effect of increasing rates of foot-point heating is the flattening of the temperature profile, which begins to show a slight depression near $s_{n}$ for $\alpha = 3.0$, indicating that the heat flux, which was previously directed toward the sun, has been locally reversed, and is now transporting thermal energy up from the base of the corona, to be dissipated near $s_{n}$.
This effect, which is critical to the development of a nonequilibrium condition, becomes even more exaggerated for $\alpha=3.45$, as depicted in the bright orange curve.
There the increased radiative losses cause an outsized reduction in the temperature near $s_{n}$, and a corresponding enhancement in the density, which falls off more slowly with height between $s_0$ and $s_{n}$ as $\alpha$ is increased.

The development of a new local minimum in $T$ near $s_{n}$, which is several $10^{5}\,\rm K$ cooler than the top of the TR, generates a large upward conductive heat flux between $s_0$ and $s_{n}$, which transports significant thermal energy into the region surrounding the new temperature minimum, where it is lost to radiation.
This upward heat flux insulates the TR from the excess heat energy that would normally be transported down from the corona, which negates the usual effect of increased foot-point heating, so that as $\alpha$ is increased beyond $\alpha\gtrsim2$ the asymptotic wind becomes marginally faster, hotter, and less dense -- exactly opposite the behavior in the case of weak foot-point heating -- even as the plasma near $s_{n}$ becomes colder, denser, and increasingly $\mathcal{R}$-dominated.
The region surrounding the temperature minimum near $s_{n}$ therefore behaves as a thermal sink, absorbing excess thermal energy from the larger coronal volume and dissipating it locally through radiation.

But within the thermal sink, the sensitivity of the temperature to increases in $\mathcal{H}_b$ -- which is compounded by the $T^{5/2}$ dependence of thermal conduction -- implies a limit to this energy dissipation pathway.
While the examples in Figure \ref{Figure::Marginal} all represent steady state solutions, it takes only a small ($<2\%$) increase in $\alpha$ from $3.45$ to $3.5$, corresponding to a foot-point heating rate of $\mathcal{H}_b = 7.0 \times 10^{-4} \, \rm erg\, cm^{-3}\, s^{-1}$, to cause the temperature within the thermal sink to drop below $T^\prime$, at which point the model emissivity $\Lambda(T<T^\prime)$ becomes a decreasing function of $T$, making the plasma susceptible to thermal runaway.

\section{Loss of Equilibrium}\label{Section::Onset}

\subsection{Response to Increased Heating}

\begin{figure}
     \centering
     \includegraphics[width = \linewidth]{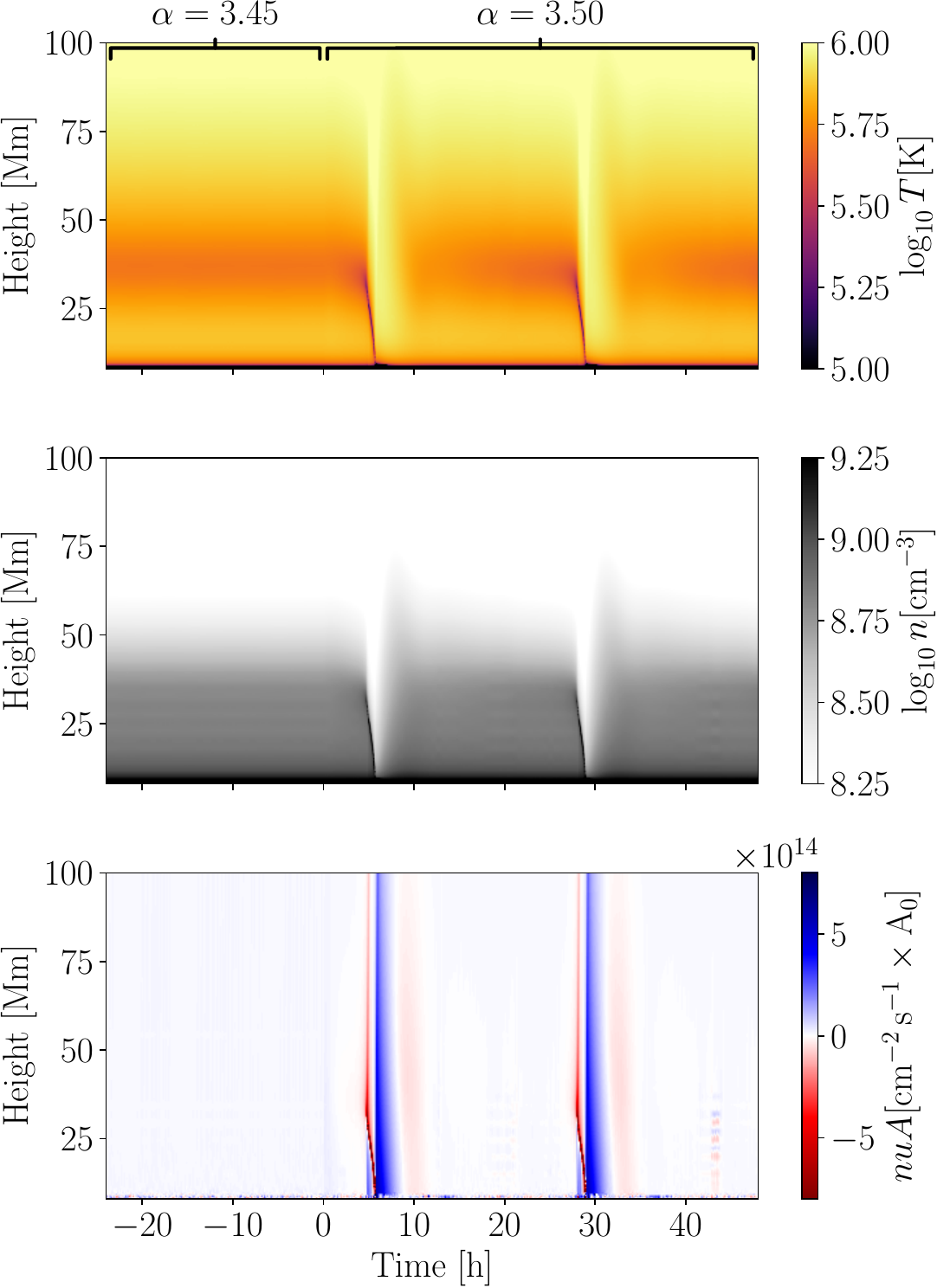}
     \caption{
     Time-distance plots of temperature (top), number density (middle), and linear mass flux (bottom) over a $72\,\rm h$ period during which the heating ratio $\alpha$ increases from $3.45$ to $3.5$, corresponding to foot-point heating rates of $\mathcal{H}_b \in \{6.9, 7.0\} \times 10^{-4}\,\rm erg\, cm^{-3}\, s^{-1}$. During the first $24\,\rm h$ ($\alpha=3.45$) the plasma is in a steady state. Following the increase in foot-point heating ($\alpha = 3.5$) at $t=0$ the plasma undergoes two TNE cycles over the subsequent $48\,\rm h$ period.}
     \label{Figure::Onset}
\end{figure}

In each of the cases presented above, the ratio of integrated heating rates was in the range $0 \le \alpha \le 3.45$, which effectively defines the upper limit to the rate of foot-point heating that the system can tolerate under steady state conditions.
When $\alpha$ is increased beyond $3.45$ the subsequent increase in density causes the temperature within the thermal sink to drop below $T^\prime$, which is the threshold temperature for thermal stability set by the model emissivity $\Lambda(T)$.
The resulting loss of equilibrium and ensuing thermal runaway are depicted in Figure \ref{Figure::Onset}.
In the figure, time increases on the horizontal axis, with $t=0$ indicating the moment when the heating is increased.
Prior to this time the plasma exhibits the same steady-state condition shown in the red curves in Figure \ref{Figure::Marginal}.
For times $t>0$ the foot-point heating is a steady $\mathcal{H}_b = 7.0 \times 10^{-4}\rm\, erg\, cm^{-3}\, s^{-1}$ (an increase of $\sim1.5\%$ over the previous value of $6.9 \times 10^{-4}\rm\, erg\, cm^{-3}\, s^{-1}$).
The subsequent evolution is shown in the time-distance maps of temperature (top panel), number density (middle panel), and linear particle flux ($n u A$, bottom panel).

Prior to the increase in heating ($t<0$) the thermal sink is visible as the dark red band in the temperature map near $s_{n}$, between $s\sim 25$ and $50\,\rm Mm$ above the solar surface (indicated on the vertical axis).
Following the increase in heating the temperature and density within the thermal sink show no obvious change at first, although close inspection reveals that they are evolving, albeit slowly.
The first indication that something has changed in a meaningful way is the reversal of the linear particle flux near the thermal sink, indicated by the faint red (inward) flux in the bottom panel of Figure \ref{Figure::Onset}.
This first appears about $2\,\rm h$ after the increase in heating, and is shortly followed by a slight increase in the rate of mass accumulation (and cooling) within the thermal sink, indicated by a small but steady darkening of the middle and upper panels near $s_{n}$.

The slow decrease in temperature and increase in density proceeds steadily until $t\approx 5\,\rm h$ after the heating increase, at which point there is an abrupt increase in density and simultaneous decrease in temperature, and a rapid increase in the downward particle flux (vertical red feature in the bottom panel).
The condensed material then falls rapidly toward the solar surface, as indicated by the dark streak in both the temperature and density maps.
Notably, while the cold/dense material is compact in its vertical extent, the reversal of the mass flux that transports this structure from the corona to the solar surface extends vertically up to a height of $s \gtrsim s_{p}$, indicating that a significant portion of the column of coronal plasma outside of the thermal sink has been bodily displaced and entrained by the falling condensate.

As soon as the falling condensate reaches the solar surface the mass flux reverses again and a strong up-flow forms, extending up to and beyond the $\sim90\,\rm Mm$ ($h_{p}$) spanned by original reversal, as the material that previously drained down from the corona is expelled from the chromosphere and transition region. 
This proceeds until about $t\sim8\,\rm h$ after the increase in heating, by which time all of the material that previously accumulated at the base of the corona is once again distributed within the larger coronal volume.
However, the replenished coronal plasma is far from equilibrium, and the plasma begins to settle as it relaxes toward a force-balanced state.

The settling is rapid at first, indicated by the red downflow in the bottom panel between $t\sim 8$ and $12\,\rm h$, and then becomes more gradual as the mass flux returns to a weak outflow condition similar to the prevailing steady-state condition prior to the increase in heating. 
During this latter phase the temperature near $s_{n}$ steadily deceases and the density increases as a new thermal sink forms, and the conditions there become conducive to thermal runaway once again.
Eventually, at around $t\sim24\,\rm h$, a second condensate develops within the thermal sink, and the process repeats.

\subsection{Effect of Flow on Cooling Rate}
\citet{Klimchuk:2019o} observed that for asymmetric coronal loops the induced siphon flow drives an enthalpy flux, which serves as a heat source that can inhibit TNE.
One could reasonably wonder why the steady outflow of the solar wind does not similarly prevent the loss of equilibrium along open field lines. 
An important difference between the simulations presented here and previous studies of asymmetric coronal loops is that increases in the foot-point heating rate cause a decrease in the velocity within the thermal sink; 
however, this alone does not preclude an increase in the enthalpy flux.
Indeed, when $\mathcal{H}_b$ is increased, $P$ increases with it, especially in the region $s_0 < s < s_{n}$ (as shown in the second panel of Figure \ref{Figure::Marginal}) so it is not obvious what effect $\alpha$ should have on the energy balance near $s_{n}$.

Previous authors have discussed the role of mass and thermal energy fluxes in the inertial (Eulerian) reference frame of the sun.
Here we will explore the effect of the flow in the co-moving (Lagrangian) frame of the fluid and consider whether a given profile of heating and cooling is consistent with a physically viable thermal evolution.
If we assume that the plasma is in a steady state, then the thermal history of a fluid parcel that emerges from above the thermal sink must be consistent with the spatial variation of the various heating and cooling terms within the sink -- that is, the time history of a fluid parcel should be equivalent to the spatial variation of the plasma properties along the flow.
If the net energy loss over the plasma's thermal history exceeds its internal energy, then the profile is understood to be unphysical, meaning that no steady-state can exist under those conditions.
A spatially viable profile of the various heating and cooling terms is, therefore, a necessary (if not sufficient) condition of a steady state.

In the Lagrangian frame the timescale for the plasma to cool under the combined influence of heating, radiation, and thermal conduction is given by the dynamic cooling time, which is the inverse of the dynamic cooling rate,
\begin{equation}
\tau_d^{-1} \equiv  -(\gamma - 1) (\mathcal{H} + \mathcal{R} + \dot{Q}_c) / P,
\end{equation}
where we have omitted the contribution from ${\rm div}(u)$ on the RHS of Eq. \eqref{Equation::temperature} as it is small compared to the other heating and cooling terms.
If $\tau_d > 0$ this indicates that the energy loss due to radiation locally exceeds the energy added by external heating and thermal conduction, so if a parcel of plasma were to experience this condition over a long enough period its internal energy would eventually go to zero; 
however, if the flow transports the plasma through the region of unsupported cooling at sufficient speed, the plasma can still exhibit a steady state.
Therefore, the reference value for $\tau_d^{-1}$ is the timescale implied by the advective derivative,  which is given in terms of the temperature length scale $L_T$ and the bulk speed of the fluid: 
\begin{equation}
\tau_u^{-1} \equiv u\, \partial_s \ln T \approx u/L_T.
\end{equation}
The cumulative effect of the locally unsupported cooling is then encoded in the integral of $\tau_d^{-1}$ over the time interval $\tau_u$.

For this calculation we will consider a region that extends from the point of lowest temperature within the thermal sink, down to the point where the heat flux switches from a cooling term to a heating term; about $10\,\rm Mm$ below the thermal sink.
This region, whose location and extent we determine at every time step, represents a column of plasma that is radiation dominated ($\mathcal{H} + \mathcal{R} < 0$) and whose temperature is maintained by an upward-directed heat flux that transports heat energy into the thermal sink from the heating-dominated region below it.
The time-integral of $\tau_d^{-1}$ can then be approximated by
\begin{equation}
    \int_0^{\tau_u} dt^\prime \tau_d^{-1} = \int_0^{L_T} ds^\prime u^{-1} \tau_d^{-1} 
    \approx \langle \tau_d^{-1} \rangle / \langle \tau_u^{-1} \rangle,
\end{equation}
where $\langle \circ \rangle$ denotes the spatial average over the specified region.
This approximation, using the ratio of averages, is less sensitive to small perturbations and, therefore, performs better as a diagnostic of quasi-static conditions than does the explicit integral.
The necessary (but not sufficient) condition of a steady state is then
\begin{equation}
\langle \tau_d^{-1} \rangle < \langle \tau_u^{-1} \rangle \longrightarrow 
\Gamma \equiv \langle \tau_d^{-1} \rangle / \langle \tau_u^{-1} \rangle < 1,
\end{equation}
where $\Gamma$ is the (spatially integrated) ``cooling factor,'' which encodes the net loss (or gain) of energy across the region of interest as a fraction of the total internal heat energy of the plasma.

\begin{figure}
    \centering
    \includegraphics[width=\linewidth]{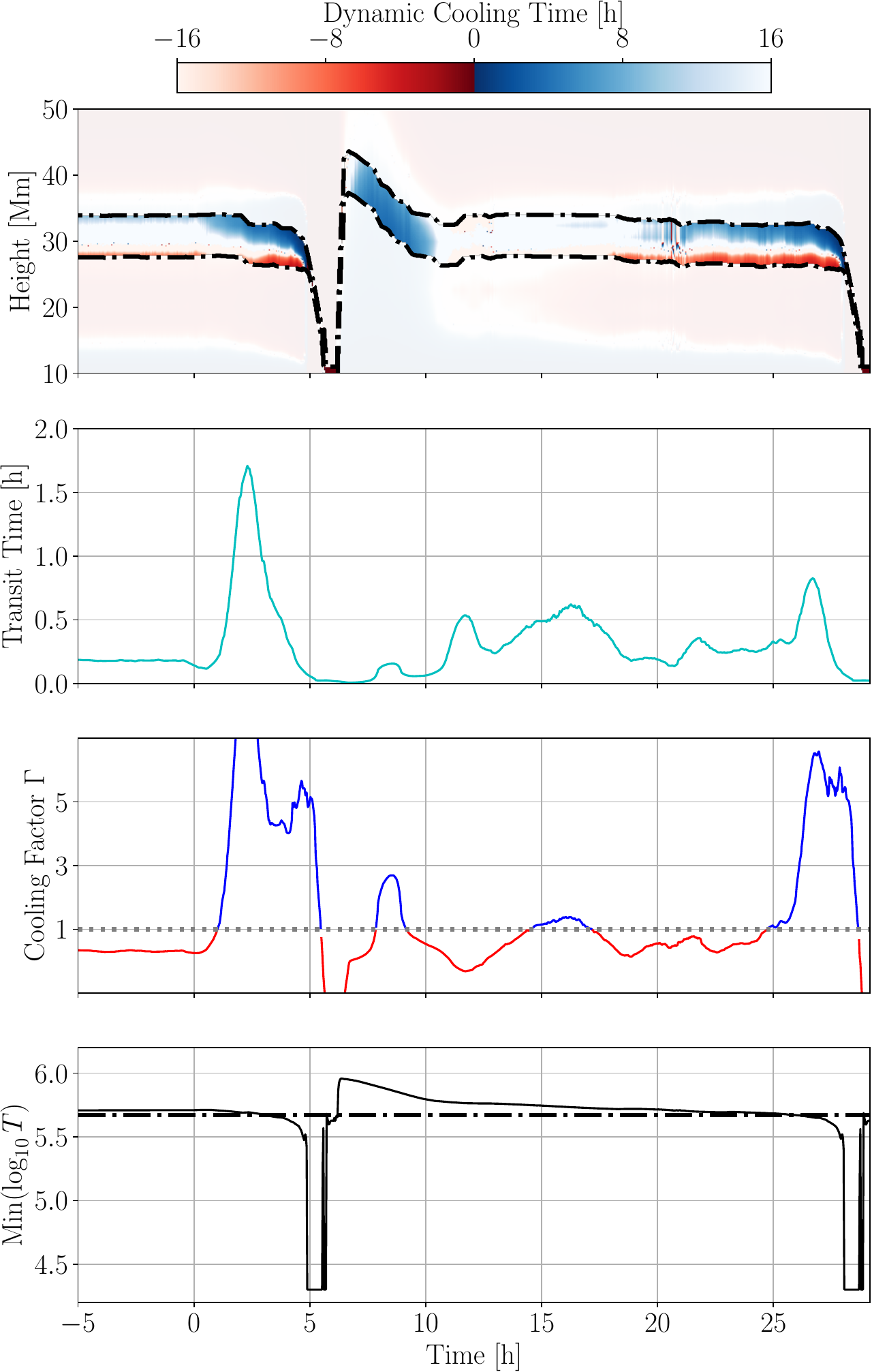}
    \caption{The dynamic cooling time, transit time, integrated cooling factor $\Gamma$, and minimum value of the temperature within the thermal sink are shown over a period of approximately $36\,\rm h$ that spans from just before the increase in heating from $\alpha = 3.45$ to $\alpha = 3.5$ at time $t=0$ until just after the end of the first full condensation cycle.}
    \label{Figure::DynamicTimescale-long}
\end{figure}

In the upper panel of Figure \ref{Figure::DynamicTimescale-long} the dynamic cooling time $\tau_d$ is plotted with respect to time in the red-blue color scale, with dark red/blue corresponding to rapid heating/cooling, while light (white) corresponds to longer (slower) heating and cooling times.
The black dash-dotted curves indicate the vertical extent of the region over which the averages are computed. 
The opacity of the $\tau_d$ color scale has been reduced outside of this region to emphasise that the values there do not contribute to $\Gamma$.
The second panel of Figure \ref{Figure::DynamicTimescale-long} represents the transit time across the thermal sink, which we approximate as $\tilde{\tau}_u = \langle u / L_T \rangle^{-1}$, where $L_T\sim 10\,\rm Mm$ is the distance between the dash-dotted curves.
The integrated cooling factor $\Gamma$ is shown in the third panel, in alternately red and blue line color indicating regions where $\langle \tau_u^{-1} \rangle$ is smaller or larger than $\langle \tau_d^{-1} \rangle$.
The bottom panel shows the temperature at the base of the thermal dip, with the critical temperature $T^\prime$ indicated by the dot-dashed reference line.

Prior to the increase in heating at $t=0$, the dynamic cooling time within the thermal sink is long, but not zero, showing that the plasma initially gains energy and then loses it as it traverses the sink, while the transit time is comparatively short, so that the net energy lost during the transit is less than the available internal thermal energy.
However, when the heating is increased ($t>0$), the plasma is immediately out of equilibrium with the heating and cooling rates, which causes $|\tau_d|$ to decrease, especially where $\tau_d > 0$, indicating rapid cooling.
Simultaneously, the flow speed through the thermal sink decreases, causing an abrupt increase in the transit time in the second panel.
The combined effect is that $\Gamma$ increases rapidly, indicating that a steady state is inadmissible and the plasma is evolving toward an overall cooler state.

This initial increase in $\Gamma$ is not itself an indication of TNE, but merely a consequence of the fact that the new heating rate (and corresponding radiative losses) are inconsistent with the equilibrium state that prevailed prior to the increase in $\mathcal{H}_b$.
On close inspection we can see that at about $t\sim 2\,\rm h$ the local structure of $\tau_d$ appears to stabilize and $\Gamma$ begins to decrease, suggesting that the plasma may be settling to a new steady state that is consistent with the new heating rate.
However, a short time later we see $\Gamma$ begin to grow again, just as $T$ drops below $T^\prime$, and soon after that the vertical extent of the thermal sink begins to narrow as plasma from the surrounding volume is drawn into it.

As the plasma within the thermal sink becomes increasingly dense, it rapidly cools and coalesces into a condensate that ultimately falls toward the solar surface and is subsumed by the model chromosphere.
This process leaves the lower corona in an under-dense state and creates an imbalance within the TR at the interface between the recently-evacuated corona and the newly-formed upper layer of the chromosphere.
Consequently, the upper layer of the chromosphere begins to evaporate as the material that previously formed the condensate is ablated and expelled back into the corona, and while this process is proceeding the cooling factor $\Gamma$ has no utility as a counterfactual test of a steady state since the plasma is clearly evolving at a rapid rate.
Only after the plasma parameters have begun to stabilize and the thermal sink is reforming, around $t\gtrsim20\,\rm h$, do we again see that the structure of the dynamic cooling rate within the thermal sink is quasi-steady, with moderate net energy loss, and the transit time is reasonably short, meaning that the flow is able to carry the plasma across the region of unsupported cooling before its internal thermal energy is entirely lost to radiation.

During this time the flow is preventing the onset of a thermal runaway, and in the Eulerian reference frame of the Sun we would say that the heat supplied by the enthalpy flux is acting to partially offset the radiative cooling.
However, as is evident in the lower panel, the temperature within the thermal sink never fully stabilizes, and as the density near $s_{n}$ slowly rises, steadily increasing the radiative energy loss, the temperature eventually drops below $T^\prime$.
Once this happens the radiative loss rate begins to increase with decreasing temperature, further reducing the dynamic cooling time, as seen in the rapid increase in $\Gamma$ at around $t=25\,\rm h$.
Then, just as in the previous instance, the plasma begins to cool rapidly within the thermal sink before coalescing and falling toward the solar surface once again.

\section{Phases of a TNE cycle}\label{Section::Phases}

During the initial evolution away from a steady state the residual effects of the abrupt (albeit small) increase in $\mathcal{H}_b$ continue to influence the plasma dynamics up until the point that the temperature within the thermal sink has dropped to the critical value of $T^\prime \simeq 0.47\,\rm MK$.
Beyond this point the subsequent evolution is a quasi-periodic limit cycle of precipitation and evaporation, which shows no meaningful variation in its behavior between subsequent cycles.
Within a given cycle the evolution can be divided into four distinct phases, each of which is governed by a dominant physical process that dictates a particular timescale for the evolution. 
We begin our enumeration of these phases following the formation of the first condensate.

\subsection{Precipitation}

\begin{figure}
    \centering
    \includegraphics[width=\linewidth]{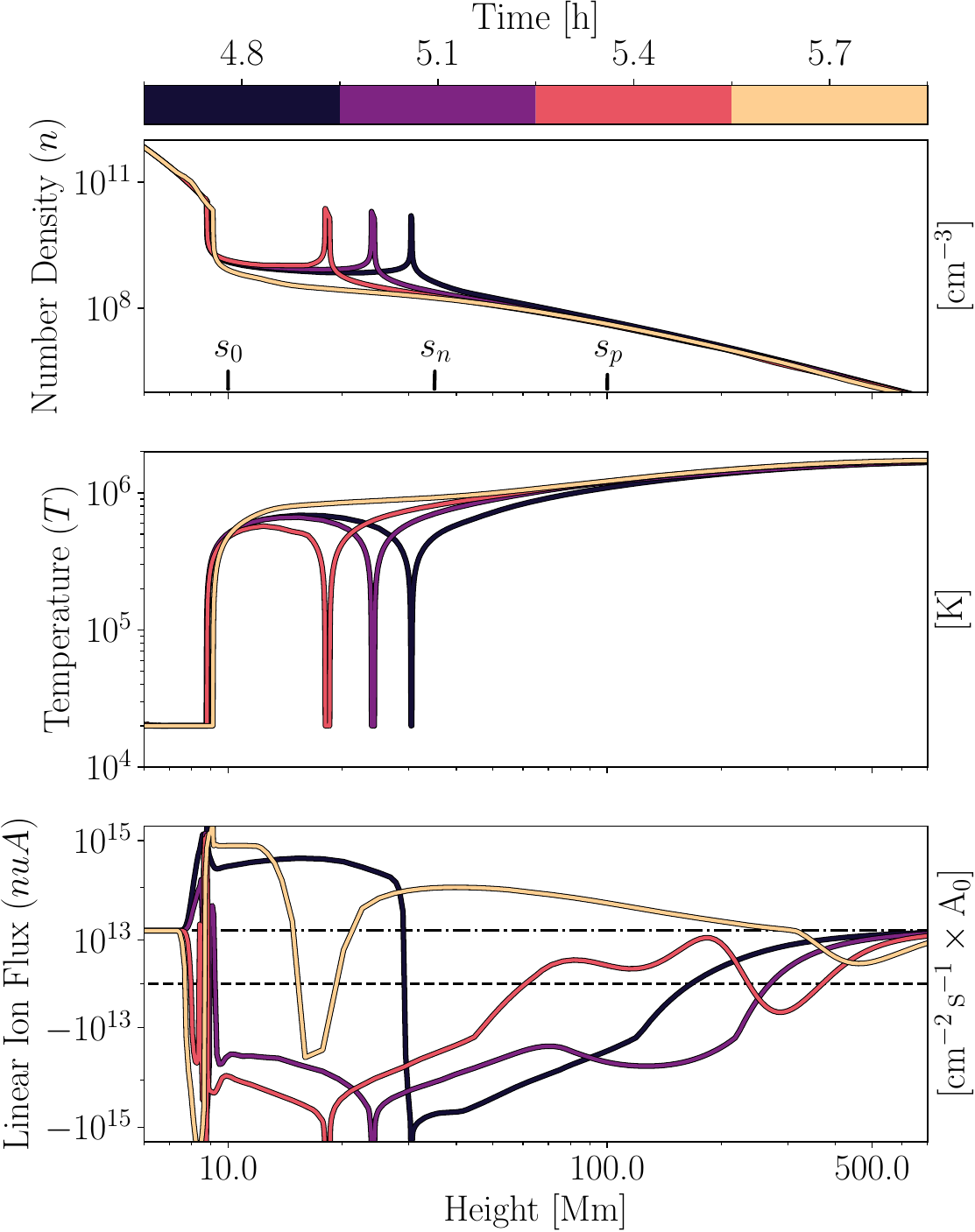}
    \caption{The time evolution of $n$, $T$, and $n u A$ are depicted during the precipitation phase, which lasts approximately $1\,\rm h$ for the case of marginal nonequilibrium, beginning with the formation of a cold condensation in the corona near $s_{n}$, and ending with the up-welling of material at the base of the corona.}
    \label{Figure::Precipitation}
\end{figure}

The evolution of the plasma following the formation of the condensate is shown in Figure \ref{Figure::Precipitation}, with the number density, temperature, and linear particle flux ($nuA$) displayed in the top, middle, and bottom panels, respectively.
The variously colored curves indicate simulation snapshots at different times during the interval from $t\approx4.8\, \rm h$ to $t\approx5.7\, \rm h$ following the increase in $\alpha$ from $3.45$ to $3.5$ at $t=0$.
The black curve corresponds to the state of the plasma at the exact moment that the temperature within the thermal sink reaches $T=T^\prime$, indicating the end of the condensation phase (described later) and the onset of the precipitation phase.

While the reduction in temperature within the thermal sink during condensation is a consequence of the imbalance between external heating, thermal conduction, and radiative cooling, the corresponding density enhancement results directly from a convergence in the flow following a (small) reduction in the plasma pressure within the sink, which draws material into the sink from above and below.
This convergent velocity profile persists into the start of the precipitation phase, with the column of plasma between the TR and the thermal sink exhibiting a strong upflow while the region above the thermal sink exhibits draining up to a height of $s \sim s_{p}$.
This reduces the density and, therefore, the pressure between the top of the TR and the developing condensate, so that the pressure gradient across the thermal sink weakens, even as the mass of the condensate increases, until the pressure force is unable to support the weight of the condensate, which begins accelerating downward toward the solar surface.

As the condensate falls the inward particle flux between $s_{n}$ and $s_{p}$ becomes weaker with time and eventually returns to an outflowing state.
This launches a small transient that propagates outward into the prevailing wind as a localized depletion in the particle flux.
Once the condensate reaches the base of the TR it is subsumed into the chromosphere, which expands upward, so that the exponentially-decaying density profile of the hydrostatic column below the TR is extended vertically to accommodate the increased mass.
Following this coalescence, a new transition region forms above the subsumed condensate, with lower density at its base, owing to reduced density at the top of the extended chromospheric column.
Meanwhile, the abrupt deceleration of the condensate at the base of the corona launches a second pressure wave that propagates upward into the downward-flowing material above it.
This causes a new outflow to develop at the base of the corona, which extends to progressively larger heights behind the upward-propagating pressure front.

The trajectory of the condensate itself is similar to ballistic free fall; however, as the condensate falls, it drives and entrains coronal material in front of and behind it, which causes its mass to increase in time and also increases the pressure differential across the condensate, both of which limit its free-fall speed.
To first order the effect of mass accumulation can be accounted for by assuming that the momentum of the falling condensate increases in response to both its acceleration, as it falls vertically toward the sun, and also the increase in mass that occurs as it falls. 
The simplified momentum equation for this behavior is described by
\begin{equation}
    \partial_t \left ( \rho A z v \right ) = (\rho A z) g_\odot
\end{equation}
where $z$ is the distance fallen, $\rho A z$ is the accumulated mass, $v=\partial_t z$ is the speed of the falling condensate, and $g_\odot = 0.274\rm\, km\, s^{-2}$ is the gravitational acceleration near the solar surface, which is positive in this reference frame.
Treating $\rho$ and $A$ as constant background quantities, this equation admits a quadratic solution
\begin{equation}
    z(t) = \frac{g_\odot}{6} t^2,
\end{equation}
which can be found by assuming a polynomial form for $z(t)$ and setting $z(t=0) = v(t=0) = 0$.
Assuming that the condensate falls a distance of $h_{n}$, the associated free-fall time is $\tau_g = \sqrt{6 h_{n}/g_\odot} \sim 23\,\rm minutes$.

The formation of the new transition region above the recently subsumed condensate and the return of the particle (mass) flux to an outflowing condition in the lower corona signal the end of the precipitation phase and the start of the ablation phase.

\subsection{Ablation} 

After absorbing the coronal condensate, the top of the chromosphere is lifted by a height of $\delta_{ch} \sim 500 \rm \, km$, while the density profile within the chromosphere continues to decay exponentially with height, owing to the isothermal and near-hydrostatic conditions below the newly-formed transition region.
As a result, the density at the base of the newly-formed TR is smaller by about a factor of two ($\sim 2$ vs. $4 \times 10^{10}{\,\rm cm}^{-3}$) compared to the previous state.
Since the radiative losses scale with the square of the density, while the heat capacity is proportional to the density, the effect of this change in density is to largely eliminate the contribution of radiation to the energy balance and substantially decrease the heating timescale, meaning that the plasma at the base of the newly-formed TR behaves as if the heating function had been abruptly increased, similar to a weak flare-heating event.
Consequently, the material at the base of the TR (i.e., the upper-most layer of the chromosphere), which previously made up the mas of the coronal condensate, is subsequently ablated and expelled back into the coronal domain whence it originated.

\begin{figure}
    \centering
    \includegraphics[width=\linewidth]{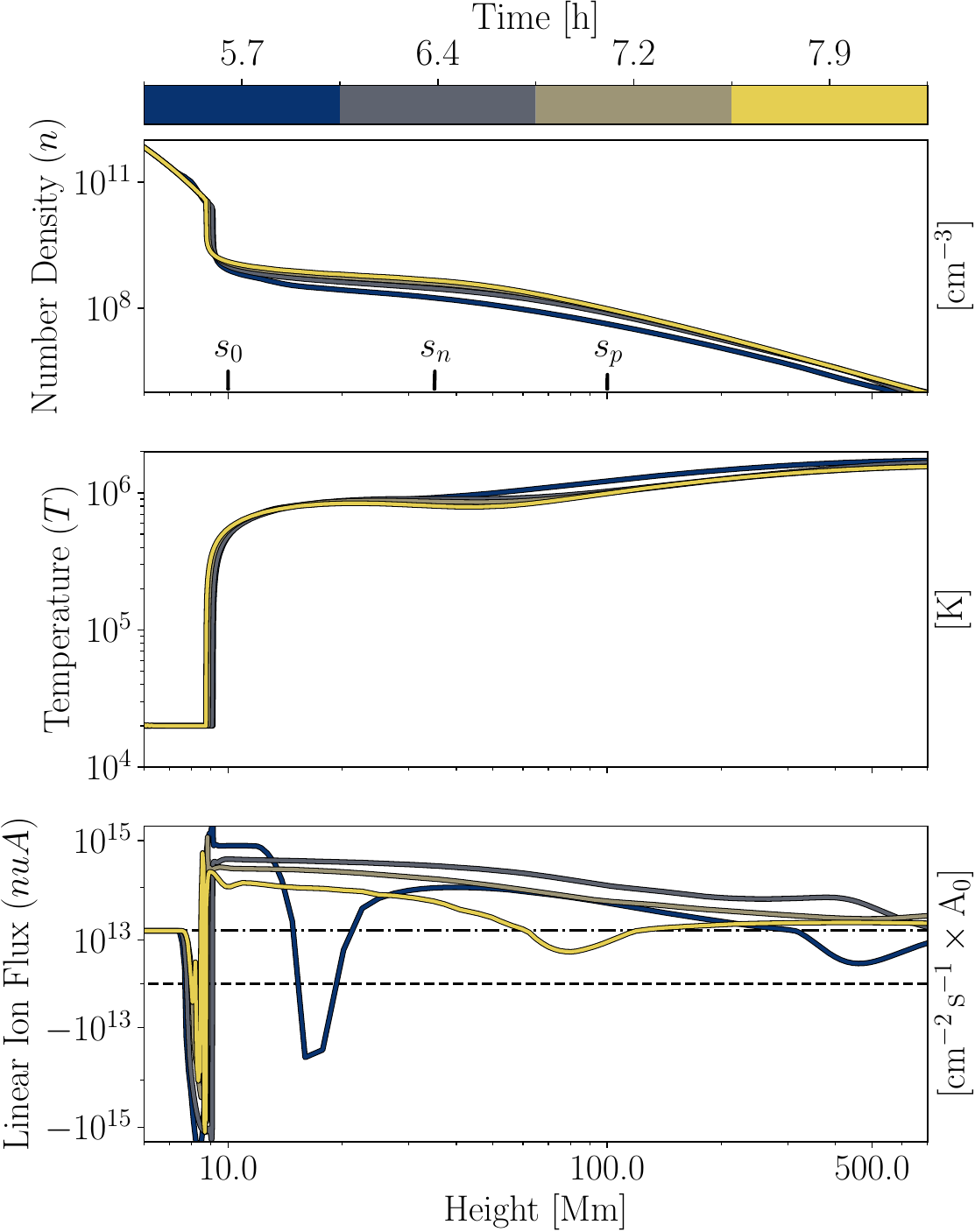}
    \caption{The time evolution of $n$, $T$, and $n u A$ are depicted during the ablation phase, which lasts approximately $2.2\,\rm h$ for the case of marginal nonequilibrium, beginning with the upwelling of material at the base of the TR and ending when all of the precipitated condensate material is expelled back into the corona and the outward mass near $s_{p}$ begins to slow.}
    \label{Figure::Ablation}
\end{figure}

This process is depicted in Figure \ref{Figure::Ablation}, which shows a time-series of snapshots from $t\approx5.7\,\rm h$ to $t \approx 7.9\,\rm h$.
Initially, there is a strong upward particle flux at the base of the corona, beneath a remnant downflow from the precipitation phase, which gives way to an outflow in the extended corona.
As time advances the remnant downflow is halted by an outward-propagating pressure front that causes the accretion to stall, and increases the particle flux through the entire column.
The pressure front travels upward at approximately the ion thermal speed so that by the time of the second snaphsot ($t\approx 6.4\,\rm h$) it has passed beyond the height where inflows were observed in the previous phase and restored the outward particle flux throughout the corona.

The ablation of chromospheric material lasts until the equivalent mass of the condensate, which previously formed the upper layer of the chromosphere, has been expelled back into the corona.
During this time the location of the TR moves steadily downward as the top of the chromosphere recedes, so that the density locally decreases at locations that were previously within the TR but which have become part of the coronal domain.
However, when referenced to the location of the interface between the chromosphere and the TR, the density at the top of the chromosphere rises steadily, as does the density within the TR and, indeed, at all heights above it. 
This process ultimately replenishes the plasma in the coronal volume and (temporarily) restores the energy balance within the TR.

The time required for the precipitated material to be returned into the corona is governed by the total mass of the condensate, divided by the mass flux of the ablated plasma through the TR, or, equivalently, the thickness of the precipitated layer divided by the speed at which the top of the chromosphere recedes during ablation.  
This can be estimated by assuming that the energy balance at the top of the TR is dominated by the thermal energy and conductive heat fluxes (radiative losses having become negligible by the momentary reduction in density), and that the conductive heat flux is, itself, powered by the integrated foot-point heating. 
Then, because the heat flux is zero at the end-points of the column (from the bottom of the TR to the local temperature peak), integrating over this volume yields
\begin{equation} 
|[\, P u\, ]|_{\rm TR} \approx (\gamma - 1) \mathcal{H}_b \ell_b,
\end{equation}
where $|[\, \circ\, ]|_{\rm TR}$ indicates the jump across the TR, from the top of the chromosphere to the local temperature peak at $s \sim s_0 + \ell_b$.
Here we have assumed that the contribution from $\mathcal{H}_g$ to the integrated heating rate is negligible within this region and that the variation in the cross-sectional area is minimal over the first few tens of $\rm Mm$ above the solar surface.

The jump in $Pu$ is dominated at the upper limit since the sound speed there is an order of magnitude larger than it is at the base of the TR and the mass flux is constant by continuity (we are assuming that the ablation proceeds as a quasi-steady process).
We can therefore approximate $P u$ to be zero at the bottom of the TR, which gives a condition on the flow at the top of the TR. 
Then, by continuity, we can relate the change in the flow speed through the TR to the density jump from the chromosphere to the corona, after which we find the speed at which the plasma flows into the TR from the underlying chromosphere to be
\begin{equation}
    u_{b} = (\gamma - 1) (\mathcal{H}_b \ell_b n_c) / (P_c n_b). 
\end{equation}

Assuming that the dense chromospheric plasma remains quasi-static, this speed represents the rate at which the interface between the chromosphere and the TR moves downward as the upper layer of the chromosphere is heated and expelled through the TR into the corona.
Taking the integrated foot-point heating to be $\mathcal{H}_b \ell_b = 4.9 \times 10^5 \,\rm erg \, s^{-1}$, the pressure at the base of the corona to be $P_c \simeq 0.2 {\rm \, erg\, cm^{-3}}$, and the density jump across the TR to be $n_c/n_b \simeq 1/100$, we have $u_b \sim 0.25 \rm \, km\, s^{-1}$.
We can then estimate the timescale of the ablation to be $\tau_a = 2 \delta_{ch} / u_{b}$, where we have assumed an average velocity of $u_{b}/2$ over a traveled distance $\delta_{ch}\simeq 500\,\rm km$ corresponding to the thickness of the precipitated layer, which we have determine by inspection.
With these values, the estimated ablation timescale is $\tau_a \approx 1.1\,\rm h$.

The ablation of chromospheric material proceeds until all of the precipitated material has been expelled into the corona, and the particle flux through the TR begins to slow, triggering a small pressure drop that propagates up into the corona, decelerating the outflow.
The signature of this deceleration is a reduction in the particle flux in the yellow curve in Figure \ref{Figure::Ablation} near $s_{p}$ at $t=7.9\,\rm h$, which marks the end of the ablation phase and the beginning of relaxation phase.

\subsection{Relaxation}

\begin{figure}
    \centering
    \includegraphics[width=\linewidth]{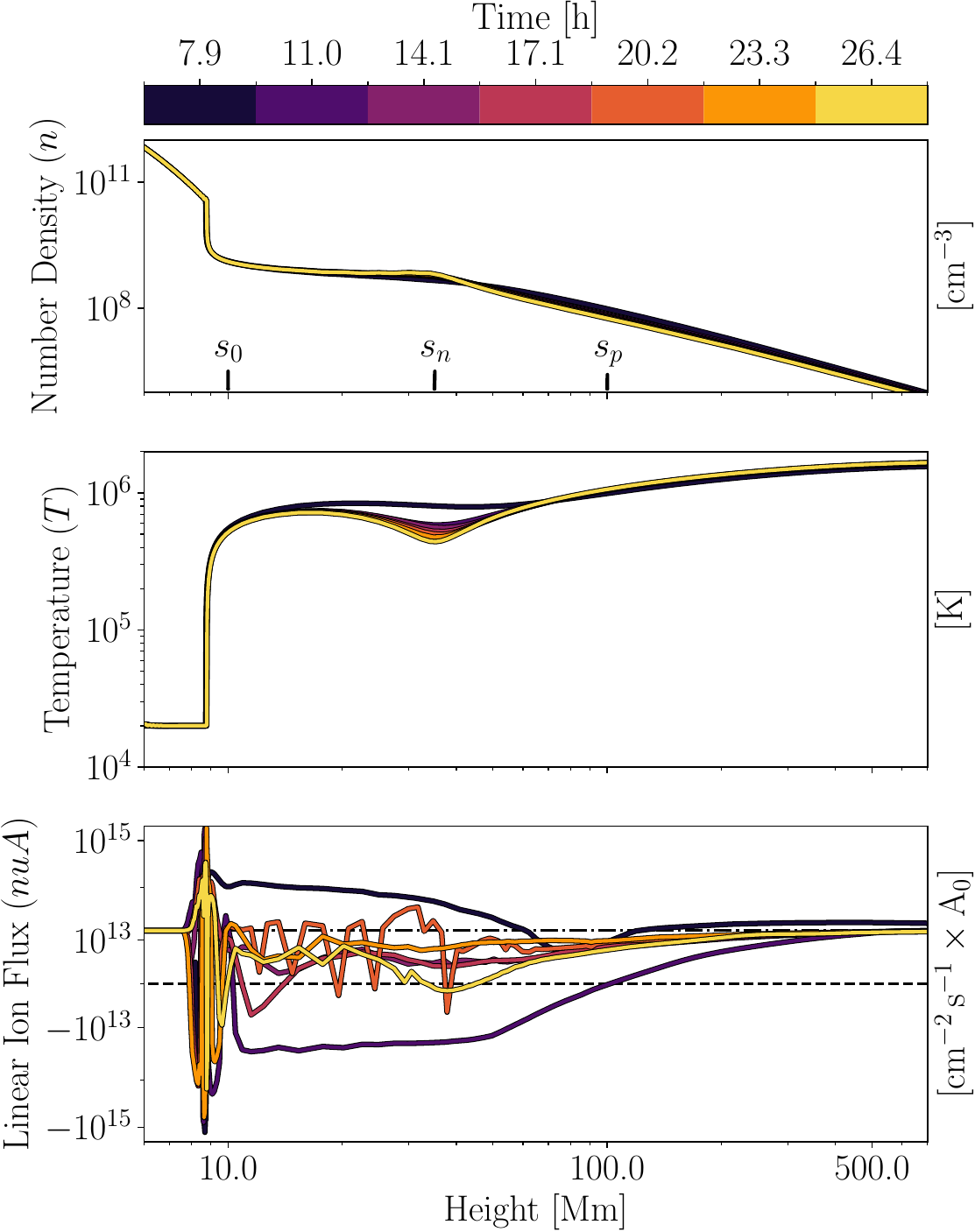}
    \caption{The time evolution of $n$, $T$, and $n u A$ are depicted during the relaxation phase, which lasts approximately $17\,\rm h$ for the case of marginal nonequilibrium, beginning with the deceleration of the outward mass flux near $s_{p}$ and ending when the mass flux develops a local convergence at $s_{n}$.}
    \label{Figure::Relaxation}
\end{figure}

The expulsion of chromospheric material through the TR during the ablation phase restores the precipitated material to the corona, but leaves substantial residual fluxes of mass and momentum that, combined with the reduction in pressure as the mass flux through the TR tapers off, implies an imbalance in the momentum and continuity equations.
These result in a sort of back-draft as the rapidly rising ablated material decelerates and then settles back into the lower corona.
This process begins with the reduction of the outward mass flux near $s_{p}$, indicated by the dip in black curve in Figure \ref{Figure::Relaxation}, and as the settling process continues the particle flux of the entire column between the TR and $s_{p}$ switches to a negative (draining) condition, as shown in the purple ($t=11.0\,\rm h$) curve in Figure \ref{Figure::Relaxation}; 

The draining-back of the plasma is short lived and by $t=14.1\,\rm h$ the particle flux becomes once again directed outward everywhere except at the very base of the corona.
At $t=17.1\,\rm h$ there is a small region of inward mass flux just above the base of the corona, which subsequently propagates outward as a pressure wave, and by $t=23.3\,\rm h$ the entire coronal domain has returned to an outward-flowing condition.
By this time the temperature and density have come to resemble the last stable equilibrium configuration depicted in Figure \ref{Figure::Marginal}, and the linear mass flux has become nearly constant, suggesting that the continuity and momentum equations are approaching a steady, force balanced state.

The initial part of this relaxation process is the most dramatic, with the mass flux of the entire lower corona undergoing rapid reversal over just a few hours.
This fast early evolution occurs because the flow speed is responding to the pressure imbalance, and the density is being redistributed by the flow, so during the early phase where the flow is fast the mass redistribution time is also fast, but as the flow speed settles to a near-steady condition the mass redistribution time becomes long.
Eventually, as the density profile begins to stabilize, the particle flux ($n u A$) between the TR and $s_{p}$ becomes increasingly uniform (except for a few spurious oscillations) and approaches the steady value set by the asymptotic solar wind, indicated by the dot-dashed line in the lower panel.
However, just as the continuity and momentum equations are coming into balance, the thermal sink is reforming and the conditions there are once again becoming conducive to thermal runaway.

Since the conditions that initially triggered the thermal runaway occurred when the plasma parameters resembled those of the previous steady state, the duration of this relaxation is approximately equal to the time required for the density to return to that (now unavailable) configuration.
The associated timescale is the mass redistribution time, which is set by the fluid transit time from the solar surface to the previous location of the thermal sink: $\tau_\rho = h_{n} / \langle u \rangle_{\rm TR}$, where $\langle u \rangle_{\rm TR} \simeq 0.5\,\rm km\, s^{-1}$ is the average flow speed in the TR and lower corona under steady conditions, which we have determined empirically from the simulation. 
Using the previously stated values of $h_{n}$ and $s_0$, the mass-redistribution time is $\tau_\rho \sim 17 \,\rm h$.

When the temperature within the thermal sink drops below $T^\prime$ the increased cooling creates a low-pressure condition that is followed by a reduction in the particle flux near the thermal sink ($s\gtrsim s_{n}$) as evidenced by the dip in the yellow curve in the bottom panel of Figure \ref{Figure::Relaxation} at $t=26.4\,\rm h$.
This marks the end of the relaxation phase and the beginning of the condensation phase.

\subsection{Condensation}

\begin{figure}
    \centering
    \includegraphics[width=\linewidth]{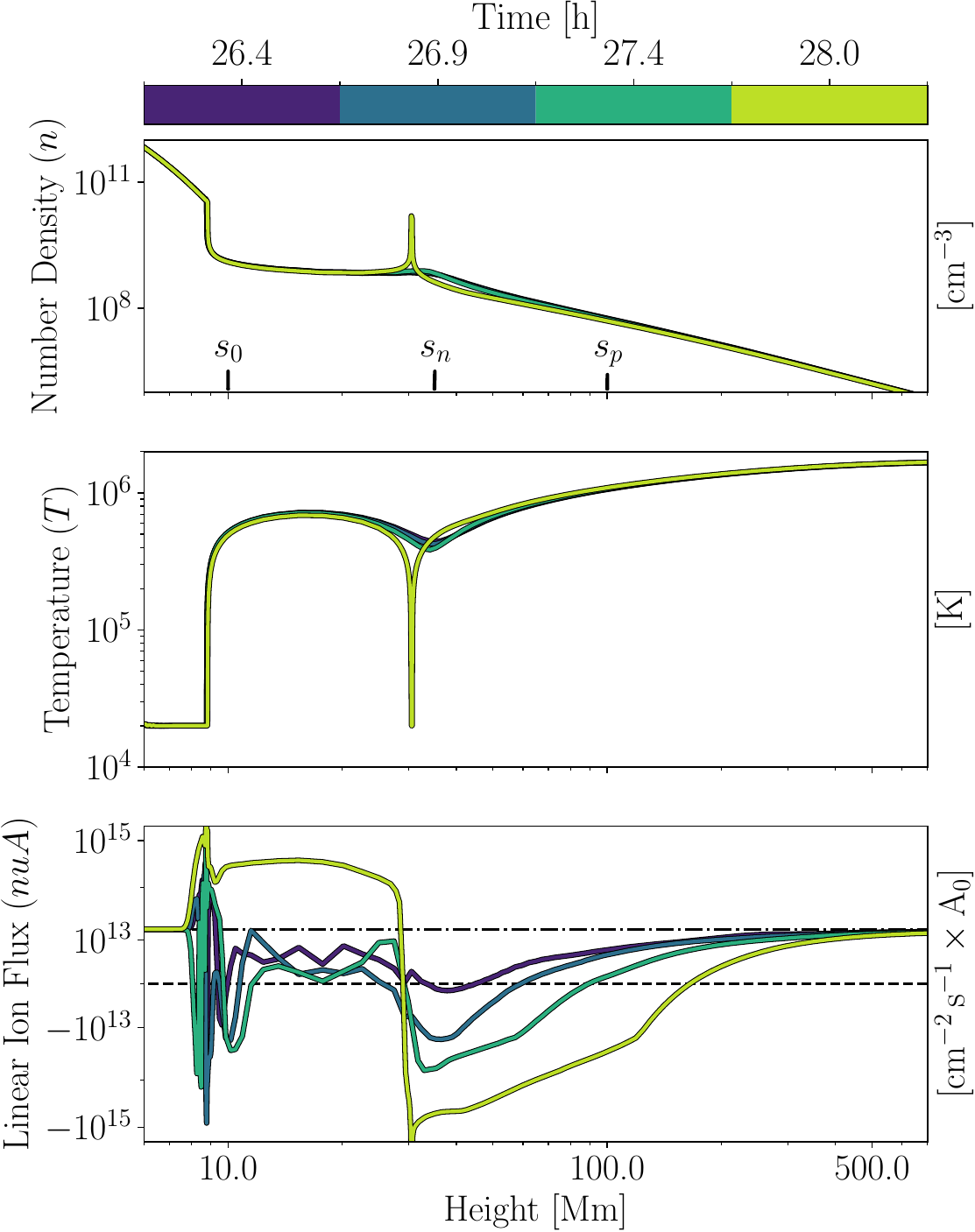}
    \caption{The time evolution of $n$, $T$, and $n u A$ are depicted during the condensation phase, which lasts approximately $1.5\,\rm h$ for the case of marginal nonequilibrium, beginning with the development of a convergent flow near $s_{n}$ and ending when the temperature near $s_{n}$ drops to chromospheric values and the trajectory of the condensate begins to accelerate rapidly downward under the influence of gravity.}
    \label{Figure::Condensation}
\end{figure}

The reduction in the particle flux near $s_{n}$ that marks the end of the relaxation phase creates a convergence in the flow across the thermal sink, which causes mass to begin accumulating there.
This is both a response to, and also an integral part of, the runaway cooling process that occurs when the temperature within the sink falls below $T^\prime$, so that $\Lambda(T)\propto T^{-1}$.
Below this threshold the radiative loss rate increases directly through the temperature dependence of $\Lambda$, so that the energy lost to radiation increases faster than the conductive heating rate as the temperature within the thermal sink decreases.
The subsequent thermal runaway and formation of the condensate are depicted in Figure \ref{Figure::Condensation}.

Once the thermal runaway has begun, the mass flux above the sink begins to decrease rapidly until the entire column above the thermal sink up to $s_{p}$ is draining downward.
Meanwhile, the mass flux below the thermal sink initially decreases, but eventually increases toward the end of the condensation phase, so that by the time the temperature within the sink has cooled to near chromospheric values the particle flux on either side of the sink is directed toward it (being an upward flux below $s_{n}$ and a downward flux between $s_{n}$ and $s_{p}$) with a magnitude that is many times larger than the asymptotic particle flux of the solar wind. 

As the mass flux within the thermal sink becomes increasingly negative its gradient also steepens in the region below the thermal sink (to the left of the temperature dip in Figure \ref{Figure::Condensation}).
This causes the density to rise rapidly until finally, when the temperature reaches $T_{\rm ch}$, the density within the sink attains a peak value of $\sim1.7\times10^{10}\,\rm cm^{-3}$. 
Interestingly, during this phase, the pressure remains comparatively smooth, varying only by a factor of two or so within the sink during the entire accumulation process, so while the reduction in pressure that results from the thermal runaway ultimately dictates the convergent character of the flow, the variation in the pressure relative to the background is relatively weak and the plasma remains in a nearly pressure-balanced (isobaric) state during the entire process. 

In order to estimate the timescale for the thermal runaway we assume that the condensation is the result of an instability that develops from an equilibrium condition, with a growth rate that depends on the details of the external heating and radiative cooling.
This assumption is inconsistent with the presumed lack of an underlying equilibrium condition in TNE; however, there is reason to believe that when the timescale of radiative cooling is shorter than the other physical timescales in the system this analysis can still be applied \citep{Klimchuk:2019d}.
Moreover, when the heating is only incrementally larger than the last stable configuration, the plasma is in a near-steady state at the end of the relaxation phase, and since every catastrophe has an associated instability at the limit of its equilibrium manifold \citep{Kliem:2014j}, it is not unreasonable that the thermal runaway that develops from this near-steady state would be described well by the growth of a thermal instability.

Following the derivation of \citet{Field:1965}, the condition of thermal instability is given in terms of the behavior condensation modes of the linearized Navier-Stokes equations (density perturbations with imaginary frequency) under isobaric (pressure balanced) conditions, which are characterized by the sensitivity of the energy loss function to changes in temperature and density.
The associated growth rate for a perturbation with wave number $k$ is given by
\begin{equation}
    \omega = - \frac{(\gamma - 1)}{\gamma P}\left ( \rho T \left ( \frac{\partial \mathscr{Q}}{\partial T}\right )_P + \kappa k^2 T \right ),
\end{equation}
where
\begin{equation}
    \mathscr{Q} = -(\mathcal{R}+\mathcal{H})/\rho
\end{equation}
is the net energy loss rate per unit mass density and
\begin{equation}
\left ( \frac{\partial \mathscr{Q}}{ \partial T} \right )_P =
    \left ( \frac{\partial \mathscr{Q}}{ \partial T} \right )_\rho - \frac{\rho}{T} \left ( \frac{\partial \mathscr{Q}}{ \partial \rho } \right )_T,
\end{equation}
is the variation in $\mathscr{Q}$ with respect to temperature at constant pressure, which depends implicitly on the density through the ideal gas law $P \propto \rho T$.
The dependence on $k^2$ reflects the conductive heating that arises from temperature variations associated with the condensation mode, which takes the form $\dot{Q}_c \propto \kappa T / \ell^2$, where $\ell = 2 \pi / k$ is the length scale of the growing condensate.

We can evaluate this expression with the specific form of $\mathcal{H}$ and $\mathcal{R}$ from section \ref{Section::Model}, where $\mathcal{H}$ is independent of both $T$ and $\rho$, while $\mathcal{R} \propto - \rho^2 T^{b}$, meaning that
\begin{equation}
    \rho T \left [ \left ( \frac{\partial \mathscr{Q}}{ \partial T} \right )_\rho - \frac{\rho}{T}\left ( \frac{\partial \mathscr{Q}}{ \partial \rho } \right )_T \right ] = (1-b) \mathcal{R} - \mathcal{H}.
\end{equation}
The growth rate is then
\begin{equation} 
\omega = \frac{(\gamma - 1)}{\gamma P}\left ( (b-1)\mathcal{R} + \mathcal{H} - \kappa T k^2 \right ),
\end{equation}
which is real and positive for wave numbers satisfying
\begin{equation}
   k^2 < \left ( (b-1) R + \mathcal{H} \right )/ \kappa T.
\end{equation}
Apparently, for wave numbers that are larger than this value (shorter length scales) thermal conduction is sufficient to transport heat in from the surrounding volume in order to prevent a thermal runaway,
but for smaller wave numbers (larger length scales), the dependence on $k^2$ means that $\dot{Q}_c$ is unable to offset $\mathcal{R}$ as the density rises and the temperature falls.
Note that for $b<1$, $(b-1)\mathcal{R}>0$ (since $R<0$) so the first term in the above expression is always positive.
The fact that the external heating $\mathcal{H}$ also enters as a positive term (contributing to the instability) is a consequence of the fact that $\mathcal{H}/\rho$ is a decreasing function of $\rho$, so its variation with respect to $\rho$ also manifests as a cooling term.

Setting $b=-1$ and assuming $T=T^\prime$ the growth rate becomes
\begin{equation} 
\omega = \frac{(\gamma - 1)}{\gamma P}\left ( \mathcal{H} - 2 \mathcal{R} - \kappa(T^\prime) T^\prime k^2 \right ),
\end{equation}
where, again, $T^\prime=10^{5.67}\,\rm K$.
Taking the density within the thermal sink to be $n = 7 \times 10^8 \, \rm cm^{-3}$, the pressure is then $P\approx0.1\,\rm erg\, cm^{-3}$. 
The conductivity is $\kappa(T^\prime) \approx 1.2 \times 10^8 \rm\, erg\, cm^{-1}\, K^{-1}\, s^{-1}$, while the radiative cooling is $\mathcal{R} = - 7 \times 10^{-5} \rm erg\, cm^{-3}\, s^{-1}$ and the external heating is $\mathcal{H}(s\sim s_{n}) \approx \mathcal{H}_g = 2\times10^{-6}\,\rm erg\, cm^{-3}\, s^{-1}$.

By inspection, the amplitude of the growing density perturbation is largest within the thermal sink and goes to zero at the base of the corona, a distance of $\sim h_{n}$ below it.
Taking this to be $1/4$ of the wavelength (the distance between the first node and antinode) of the fastest growing mode, the corresponding wave number is $k_c = \pi / (2\, h_{n})$, for which the corresponding growth rate is $\omega_c \approx 4.8\times10^{-4}\,\rm s^{-1}$. 
Then, since $\omega_c^{-1}$ represents the e-folding time of the growing mode, the expected time for the density to increase by a factor of 25 (corresponding to the observed density of $\approx 1.7\times10^{10}\,\rm cm^{-3}$ at the end of the condensation phase) is $\tau_c = \ln(25) / \omega_c \sim 1.9\,\rm h$.

The thermal runaway proceeds until the temperature within the condensate reaches a value of $T_{\rm ch}$, corresponding to a density in excess of $10^{10}\,\rm cm^{-3}$, at which point $\mathcal{R}\rightarrow0$, emulating the balance between background heating and the comparatively weak radiative losses at chromospheric temperatures.
The presence of this cold, dense plasma near $s_{n}$ marks the end of the condensation phase and the beginning of the precipitation phase.

\section{Variation in TNE Cycle Period}\label{Section::Periodicity}

In the previous section we discussed the duration of the various individual phases of a TNE cycle with a foot-point heating rate that was only slightly larger than the adjacent steady-state solution. In that case the cycle duration was primarily determined by the relaxation time required for the flow speed to become sufficiently small that the transit time exceeds the dynamic cooling time within the thermal sink, so that the temperature there drops below $T^\prime$ and the plasma undergoes thermal runaway.
In the generic case, the foot-point heating is farther from the marginal value, which reduces the duration of the various phases discussed above, each of which depends (directly or indirectly) on the heating rate.

\begin{figure}[ht]
\centering
\includegraphics[width=\linewidth]{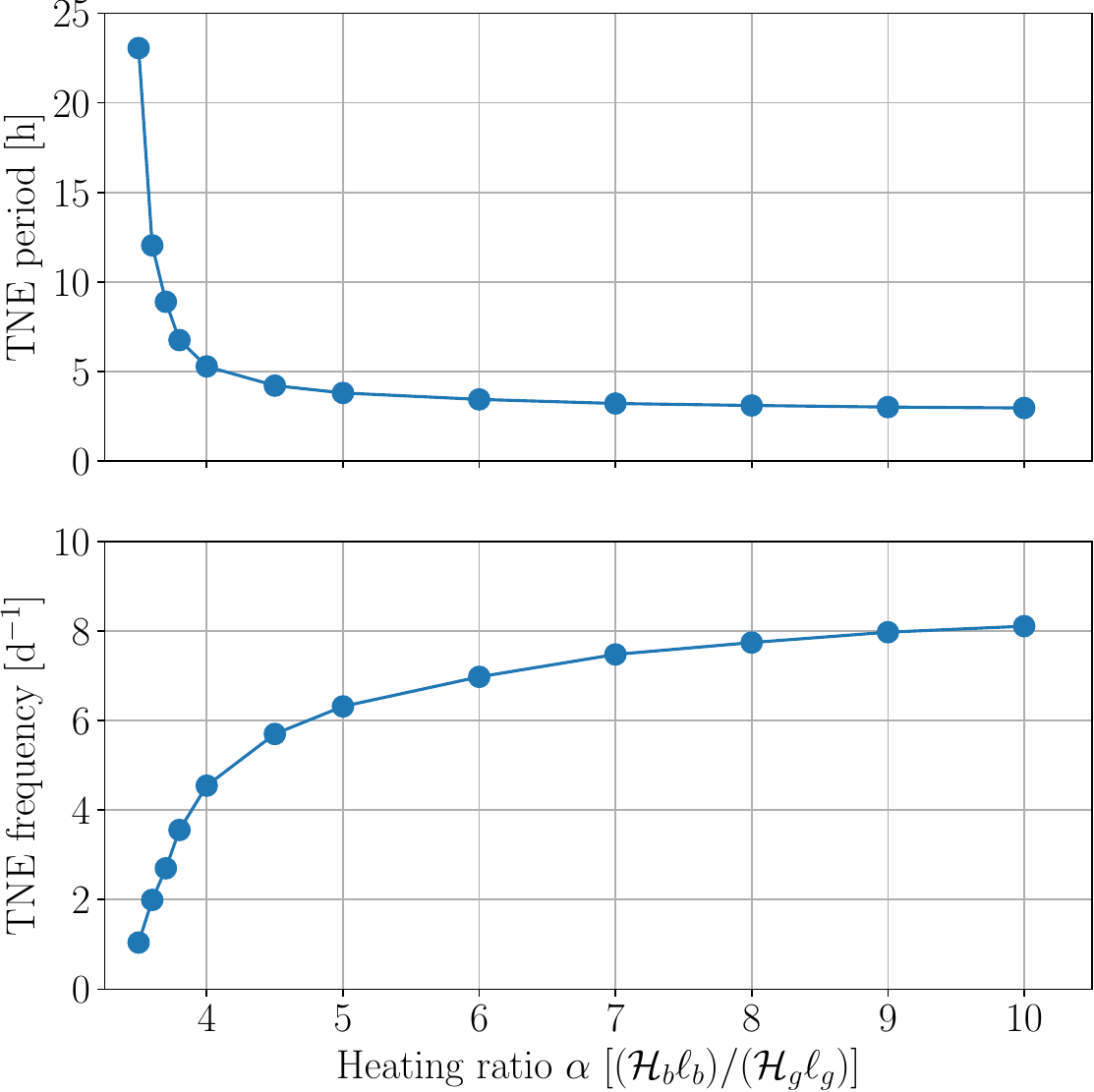}
\caption{Dependence of TNE-induced cycle duration (and frequency) on the heating ratio $\alpha$. These values represent the average cycle duration over a $56\,\rm h$ simulation interval for each value of $\alpha$, measured from the moment of condensation.}
\label{Figure::Frequency}
\end{figure}

To explore this dependence, we performed additional simulations with the same global heating rate and incrementally larger values of the foot-point heating, so that the heating ratio varied over a range of values from $3.5 \le \alpha \le 10$.
The cycle duration for each value of $\alpha$ is shown in Figure \ref{Figure::Frequency}.
Within a given cycle, the duration of each phase depends on the different physical processes that drive the evolution, each of which responds differently to changes in $\mathcal{H}_b$.

Counter-intuitively, the relaxation phase is the most sensitive to the foot-point heating, despite the fact that the timescale estimated above depends only on the flow speed and the spatial scale of the heating; however, as the heating rate, and therefore the radiative cooling, become stronger, the onset of thermal runaway occurs in the presence of larger values of $Pu$ and the relaxation process is effectively short-circuited.
Consequently, as $\mathcal{H}_b$ gets larger the duration of the relaxation phase effectively goes to zero, so that the thermal runaway occurs almost immediately after the completion of the ablation phase.
The condensation phase also depends strongly on the foot-point heating rate, which directly sets the growth rate of the thermal runaway, and indeed the observed cooling time during condensation becomes insignificant as $\mathcal{H}_b$ is increased.

The timescale for the ablation phase nominally depends on $\mathcal{H}_b^{-1}$, as this dictates the mass flux that is required to expel the subsumed material back into the corona;
however, ambient conditions of the plasma in the TR also depend on $\mathcal{H}_b$, in a way that tends to weaken this dependence, so that in practice it appears that the duration of the ablation phase tends asymptotically to $\sim 2.5\,\rm h$.
The precipitation phase is also largely unaffected by changes in the heating, as the ballistic free-fall time depends primarily on the height at which the condensate forms, and while this does get smaller for larger values of the foot-point heating, the effect is small.

Consequently, where the cycle period of the marginal case ($\alpha=3.45$) is about $24\,\rm h$, it takes only a small increase in the heating to $\alpha = 4$ to reduce the cycle period to a mere $5\,\rm h$, largely due to the reduced duration of the relaxation phase.
The cycle period continues to drop with each increase in the foot-point heating rate, but appears to approach an asymptote of $\sim 3\,\rm h$ for heating ratios of $\alpha \simeq 10$ or more.
This cycle duration is largely aligned with the results of \citet{Johnston:2019}, who found similar values (approximately $3\,\rm h$) in simulations of TNE on closed coronal loops.
In all likelihood there is no true lower bound to this value as further increases in the foot-point heating cause the condensate to form lower in the corona, which ultimately reduces both the spatial and temporal scales of the evolution; 
however, for realistic values of the $\mathcal{H}_b$, and subject to the other parameters implicit in the model, this appears to be a practical lower limit on the duration of TNE cycles.

\section{Heliospheric Signatures}\label{Section::Signatures}

\begin{figure*}[ht]
    \centering
    \includegraphics[width=0.7\linewidth]{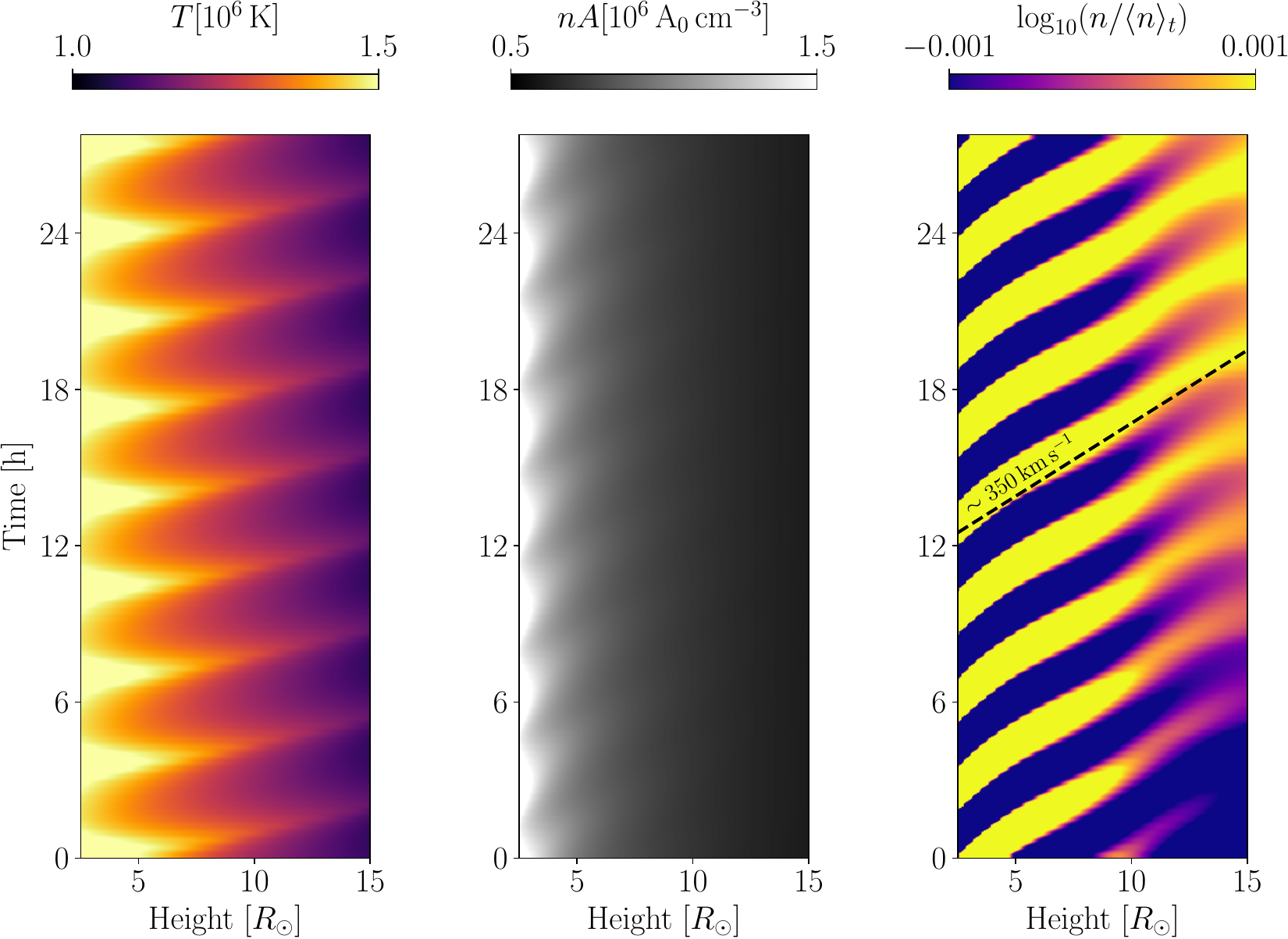}
    \caption{A time-distance plot of temperature (left), linear particle density (center), and $\log_{10}$ of the normalized particle density (right) for an interval of $10^5\,\rm s$ ($\sim 27\,\rm hr$) following an increase in the heating ratio to $\alpha = \mathcal{H}_b\ell_b/\mathcal{H}_g\ell_g = 6$. 
    Outward propagating density fluctuations occur in phase with the condensation cycles at regular intervals of $\sim3.5\,\rm h$.
    The striped pattern in the center and right panels closely matches the average characteristic speed of acoustic waves ($u+c_s$), which ranges from $250 - 450 \,\rm km \,s^{-1}$, with an average value of $\sim 350\,\rm km\,s^{-1}$ (dashed line).
    The electron thermal speed is much higher than the acoustic speed ($v_e\sim 20\times c_s$), as reflected in the slope of the bright horizontal streaks in the left panel}.
    \label{Figure::SyntheticEmission}
\end{figure*}

While the main observables of TNE cycles are likely to be found in cool emission from the lower corona, the cyclic variations in temperature, density, and flow speed near the thermal sink generate secondary signatures that are transported into the heliosphere, where they should be detectable in white light coronagraph images. 
In the left and center panels of Figure \ref{Figure::SyntheticEmission} the temperature ($T$) and linear number density ($n A$) are shown in time-distance plots.
The spatial extent (horizontal axis) of these plots spans from $2.5\,R_\odot$ to $15\,R_\odot$, which roughly corresponds to the \emph{STEREO} SECCHI/COR2 field of view.
In the right panel, the number density $n$ is shown as a fraction of the time-averaged number density $\langle n \rangle_t$ in $\log_{10}$ scale, which saturates at $\log_{10}(n/\langle n \rangle_t) = \pm 0.001$, emphasising the faint structures in the central panel.

The near-linear bright yellow and dark blue lanes in the right-most panel of Figure \ref{Figure::SyntheticEmission} show the speed of density perturbations as they propagate into the heliosphere.
The acoustic wave speed $c_s = \sqrt{\gamma P / \rho} \sim 200 \, \rm km \,s^{-1}$ over this region, owing to the nearly-constant temperature of $T\sim 1.5\,\rm MK$, while the flow speed varies between $50$ and $250\,\rm km\,s^{-1}$, with an average value of $\sim 150\,\rm km\,s^{-1}$.
The reference line, showing a representative speed of $350\,\rm km\,s^{-1}$, corresponds to the average value of $u+c_s$ over the corresponding spatial interval.
Comparing the temperature and density panels, the heat fronts associated with these compressive outflows travel much faster than the density fluctuations, near the electron thermal speed.

Periodic fluctuations in white-light observations have been observed by \citet{Viall:2015}, who found density enhancements that propagated outward at speeds between $100$ and $250\,\rm km\, s^{-1}$ with a cadence of $\sim 1-2\,\rm hours.$
The timing of these periodic density enhancements is similar to the features in Figure \ref{Figure::SyntheticEmission}, which exhibit density enhancements with periods of $\sim 3 - 4\,\rm hours$;
however, the propagation speed of the density fluctuations in Figure \ref{Figure::SyntheticEmission} is considerably faster than the observations of \citet{Viall:2015}, which appear to be transported passively by the outflowing solar wind.
This suggests that the features observed by \cite{Viall:2015} are isobaric (pressure balanced) disturbances, which are different than the fluctuations shown here.
Nonetheless, these structures may be closely related. 

Allowing for mode conversion, as should be ubiquitous within the $\beta=1$ layer surrounding the null-points of coronal pseudostreamers and related structures, it is very likely that the acoustic disturbances reported here could excite isentropic waves (pressure balanced density enhancements), which are transported passively with the flow as non-interacting density perturbations.
It is plausible, therefore, that the density perturbations that result from TNE cycles in the lower corona could be the seeds of periodic density structures that emanate from coronal streamers and pseudostreamers.
Moreover, these oscillations could also excite magnetic reconnection, which is inextricably connected to oscillations in the coronal magnetic field and plasma \citep[see, e.g.,][]{Thurgood:2019}, and could therefore be a secondary driver of interchange reconnection near coronal null points and separators.

\begin{figure*}
    \centering
    \includegraphics[width=0.49\linewidth]{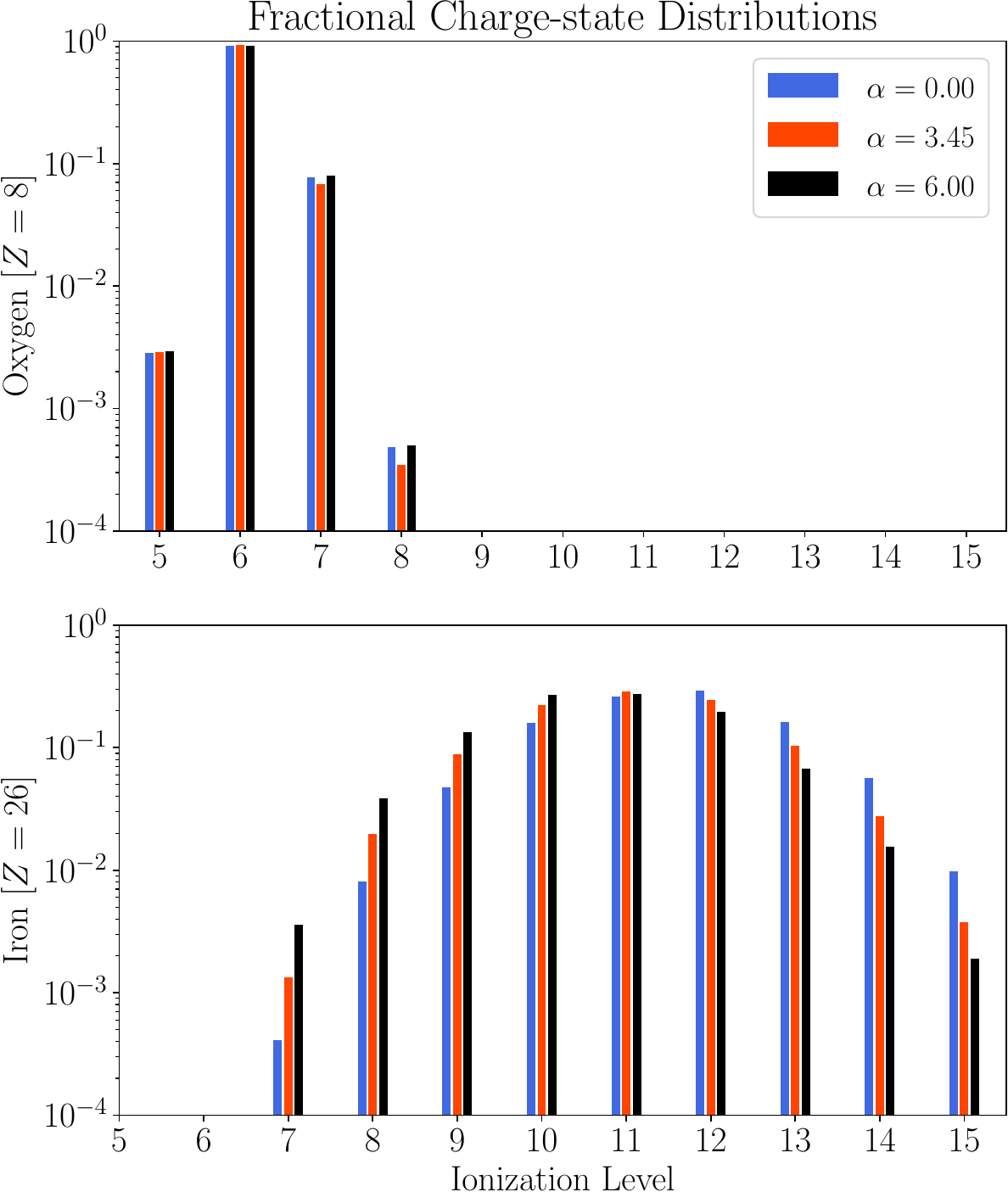}
    \includegraphics[width=0.49\linewidth]{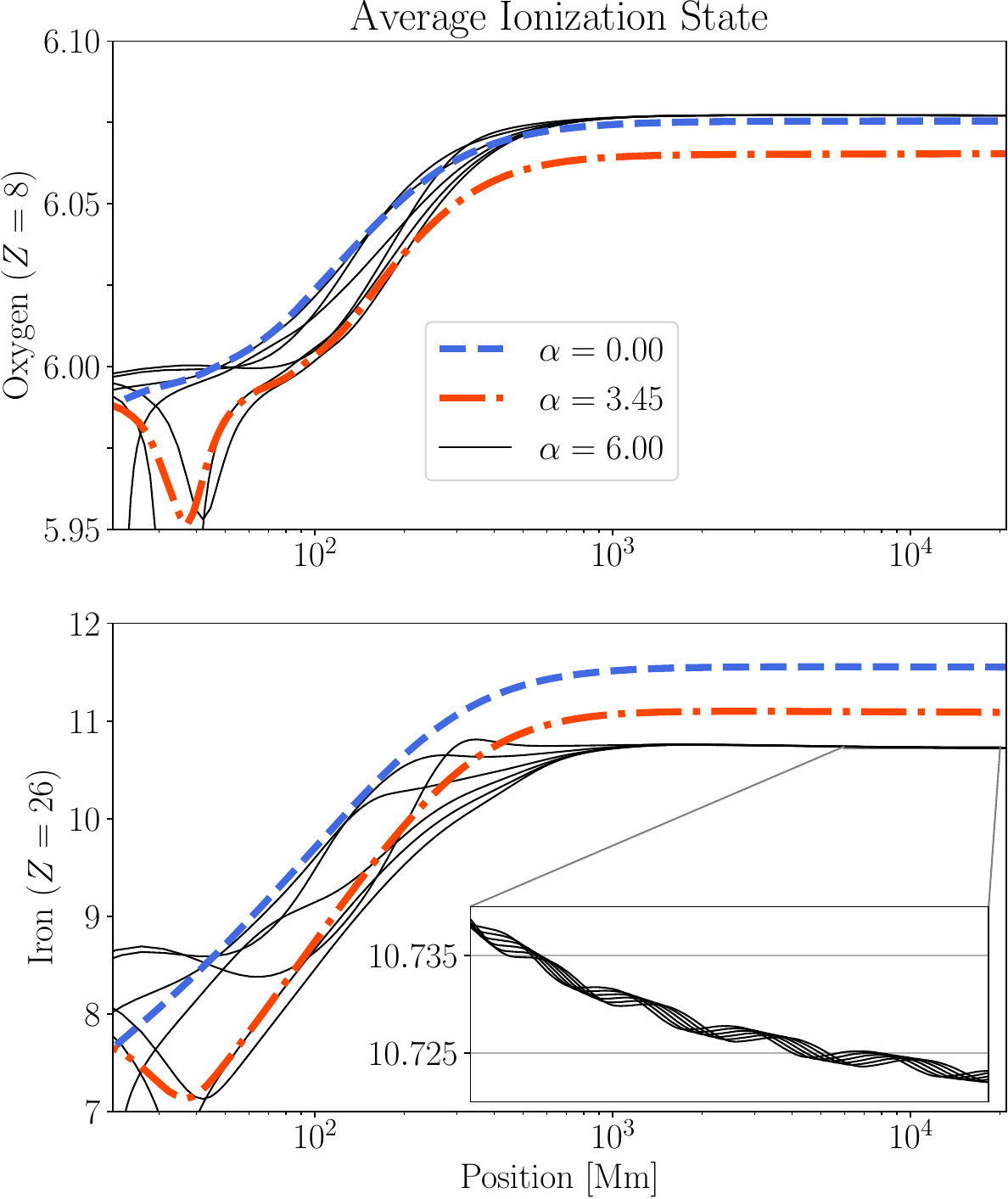}
    \caption{The charge-state distributions (left) at $r=21\,R_\odot$ ($s \sim1.5\times10^{4}\,\rm Mm$) and average charge-state (right) of Oxygen ($Z=8$) and Iron ($Z=26$) as functions of height for three characteristic values of the heating, indicated in blue, red, and black. On the right, the dashed blue and dash-dotted red curves are steady-state solutions, while the solid black curves are successive snapshots separated by $\sim 30\,\rm minutes$ during the ablation and relaxation phases of a typical TNE cycle.}
    \label{Figure::Ionization}
\end{figure*}

Another important diagnostic of the heliospheric wind is the average ionization state of trace elements, which have been used to characterize the thermal history of the plasma by such authors as \citet{vonSteiger:2011} and \citet{Szente:2022}, among others.
For the sake of computational efficiency we have chosen two representative species, Oxygen (O), and Iron (Fe), for which we calculated the time-dependent nonequilibrium ionization populations for a representative sample of heating rates, as shown in Figure \ref{Figure::Ionization}.
In the figure, the distribution of ionization states $f^Z_i$, and the average ionization level $\langle i \rangle_Z = \sum_{i=0}^{z} i f^Z_i$, are shown for both species ($Z=8$ for Oxygen and $Z=26$ for Iron), across a range of values for the heating ratio, with blue corresponding to the case of $\alpha=0$, red corresponding to the marginal equilibrium case ($\alpha=3.45$), and black representing a typical TNE condition resulting from strong heating ($\alpha=6.0$). 
When the heating ratio is increased from $\alpha=0$ to $\alpha=3.45$, the populations of ${\rm O}^{7+}$ and ${\rm O}^{8+}$ are both reduced, as are the populations of all ${\rm Fe}$ ionization levels above ${\rm Fe}^{11+}$, while those below ${\rm Fe}^{11+}$ are enhanced.
This is consistent with a reduction of the temperature near the freeze-in height, as is also visible in the spatial structure of the average ionization state, which is reduced at all heights above the thermal sink.

When $\mathcal{H}_b$ is further increased to $\alpha=6$, and the plasma begins to undergo TNE cycles, the intermittent reversal of the flow near the freeze-in height has a different effect on the ionization populations of O ($Z=8$) than it does on those of Fe ($Z=26$).
In both cases there are time-dependent fluctuations in the lower corona that largely disappear by the time the plasma gets to a height of about $R_\odot$.
Beyond that height the net effect on the ionization states of O is to negate any signature of the reduction in temperature, causing the average ionization state (and, indeed, the individual populations of O$^{7+}$ and O$^{8+}$) to return to the baseline value (similar to the $\alpha=0$ case), so that the measured ionization state in the heliosphere belies the large foot-point heating rate and the presence of a nonequilibrium condition in the lower corona.

For Fe, on the other hand, the increase in heating drives the average ionization state lower still, further enhancing (reducing) the populations below (above) ${\rm Fe}^{11+}$.
Moreover, the cyclic variation in temperature remains visible as a small time-dependent fluctuation of the average ionization state in the heliosphere, which is carried along by the outflowing wind.
Consequently, when looking at the time-dependent charge-state distribution of Fe, it is immediately clear that the temperature in the lower corona has been further reduced by the additional foot-point heating, and also that the conditions of the lower corona are not steady in time, since the charge-state distribution in the heliosphere is a direct reflection of conditions in the lower corona.
Additionally, the fact that Fe and O show conflicting trends in their asymptotic charge state distributions when the speed of the outflow in the lower corona is unsteady in time is suggestive of how future diagnostics may be developed by looking for inconsistencies in the temperatures implied by the charge-state distribution of heavy ions in the heliosphere.

\section{Discussion}\label{Section::Discussion}

The simulations presented here demonstrate the requisite conditions for TNE to occur on open magnetic field lines in the lower corona, in the regions whence the solar wind originates.
As in closed magnetic domains, it appears that the primary requirement is strong foot-point heating, which increases the density of the plasma in the lower corona, creating a radiation-dominated region that cannot be sustained by external heating or thermal conduction.
Furthermore, because the solar wind is constrained by continuity, local enhancements in the plasma density necessitate a reduction in the local wind speed (for a given mass flux), which means that for increasingly large foot-point heating rates the enthalpy flux transports proportionately less energy into the radiation-dominated region.
Consequently, while siphon flows along closed field lines have been shown to inhibit the onset of TNE \citep{Klimchuk:2019o}, the ambient outflow of the solar wind does not appear to be an inhibitor of TNE along open field lines.

Additionally, because the solar wind requires large-scale (global) heating to accelerate it beyond the sonic point, the presence of strong foot-point heating implies a local minimum in the external heating per particle, and it is in the neighborhood of this local minimum that the plasma becomes susceptible to thermal runaway.
Therefore, unlike TNE on closed field lines, which typically forms near the loop apex, the formation height of the condensate on open field lines is a direct reflection of the scale height of the foot-point heating rate.
Consequently, where condensates are observed along open field lines in the solar corona, their formation height may serve as a direct indicator of the stratification of the heating.
Similarly, any diagnostic that could be developed for inferring the formation height of coronal condensates based on the charge state distribution of heavy ions in the heliosphere would, by extension, also serve as a diagnostic for the heating scale height in the lower corona. 

Based on these arguments, it seems likely that the external heating in the interior of quiescent coronal hole regions is not sufficiently stratified to cause coronal rain to form there, as there are no known observations of condensates falling along the nearly-radial magnetic field lines that emanate from those regions.
Likewise, while the heating profiles employed here are similar to previous efforts to model coronal plumes, and give rise to similar temperature and density distributions \citep[see, e.g., ][]{Grappin:2011, Wilhelm:2011, Poletto:2015}, the fact that we do not observe coronal rain in those structures suggests that some other physical consideration may be at work there.
Conversely, where some authors \citep[e.g., ][]{Mason:2019, Li:2020} have reported observations of coronal rain forming along the separatrix surface of coronal null points and, in a few cases, along their open spines, this may suggest (1) that the external heating near these coronal magnetic null points is sufficiently stratified to create a local temperature inversion, as needed for the nonequilibrium condition to develop, and (2) that the external heating is stratified on the same spatial scale as the magnetic field itself, implying that the magnetic field is instrumental in determining the external heating rate.

In our simulations the length scale of the foot-point heating is a parameter of the model; however, 
in a physics-based heating model the spatial scale should emerge naturally from the spatial distribution of the plasma and the magnetic field.
One model that is particularly appealing in this regard is the wave-turbulence-driven (WTD) heating model of \citet{Downs:2016}, which posits that counter-propagating Alfv\'en waves accelerate and heat the plasma through reflection and turbulent dissipation.
In their model, the wave amplitudes evolve in both space and time based on the local Alfv\'en speed, the plasma velocity, and the interaction of counter-propagating modes.
Consequently, under steady-state conditions the heating rate derived from the WTD model naturally scales with the magnetic field strength \citep[$\mathcal{H} \propto B^{3/2}$ according to ][]{Downs:2016}, meaning that their model might naturally explain the occurrence of coronal rain near coronal null points, where the spatial scale of the magnetic field is dictated by the size of the corresponding separatrix dome.
It is also worth mentioning that \citet{Downs:2016} reported TNE cycles in 1D loops that were heated with their WTD model when the cross sectional area of the loop (which scales inversely with the magnetic field) was stratified on scales similar to the length scale of the foot-point heating rate employed here.

Another benefit of a self-consistent wave-heating model is that it may shed light on the different heavy ion populations that are found in the fast and slow solar wind.
The fractionation of certain elements with high First Ionization Potentials (so called high-FIP elements) is thought to be driven by the ponderomotive force, which is affected by the differing wave-energy spectra along open and closed magnetic fields \citep{Laming:2019}.
We propose that during the precipitation phase the falling condensate will act as a reflector of Alfv\'en waves owing to inverse dependence of the Alfv\'en speed on the plasma density.
This could allow the region between the chromosphere and the falling condensate to behave as a resonant cavity, which will mimic the conditions of a closed magnetic field line during that short period of time before the condensate is expelled back into the corona.
If this mechanism proves to be effective in fractionating high-FIP ions at the base of the corona, it could explain the variations in abundances between fast and slow solar wind streams, without the need to appeal to magnetic reconnection as previous authors have done \citep[see, e.g.,][and others]{Antiochos:2011.732, Higginson:2017a, Scott:2019a,Aslanyan:2021a}.

\rev{In this work we have employed the simplified model emissivity suggested by \cite{Klimchuk:2008}, which approximates the emissivity as a piece-wise continous collection of power-law profiles that span the range of chromospheric and coronal temperatures. 
We have explored the effects of including more sophistocated emissivity profiles, and we find little qualitative difference apart from small changes in the specific heating rates required to trigger the nonequilibrium condition and slight differences in the temperature at which the thermal runaway occurs. 
We have also explored the effects of allowing the temperatures of the ions and electrons to become decoupled in the low density region of the middle and upper corona.
There we find that the necessary conditions for TNE to occur are unchanged, as are the timescales that govern the various phases of the TNE limit cycles; 
however, the two-fluid energy equation introduces an additional dynamic process wherin the density and temperature undergo coherent oscillations within and below the thermal sink, which precede and eventually trigger the thermal runaway.
The presence of these oscillations suggests that the dispersion relationship for ion-acoustic waves subject to thermal conduction and radiative cooling is more complicated for a two-fluid plasma than it is when the ions and electrons share a single temperature.}

In future work we plan to incorporate a self-consistent and physically motivated heating model, such as the WTD model developed by \citet{Downs:2016}, as well as a more realistic model for the structure of the chromosphere and the details of the radiative losses, which can be calculated directly from the ionic charge-state distribution rather than \rev{a temperature-dependent emissivity profile that depends on the assumption of equilibrium ionization}.
We also intend to explore more complex magnetic field profiles, with parameterized cross-sectional areas and field-aligned gravitational forces that reflect the conditions near coronal null points and pseudostreamers.
In this way we can extend the results reported here to more realistic and relevant coronal configurations so as to better understand the conditions that lead to a state of thermal nonequilibrium, both along closed and open magnetic field lines, and to determine whether the dynamics observed within our simulations can be employed as diagnostics of the conditions in the solar corona and heliosphere.

\section{Acknowledgements}\label{Section::Acknowledgements}
This work was supported by the Office of Naval Research 6.1 Program, NASA PSP/WISPR grant NNG11EK11I, and by NASA Heliophysics Supporting Research grants under ROSES NNH20ZDA001N ``Investigating the Influence of Coronal Magnetic Geometry on the Acceleration of the Solar Wind'' (Science PI: RBS) and ROSES NNH19ZDA001N ``Explaining the Formation of Coronal Rain in Solar Flares'' (PI: JWR).
Additional support was provided by the International Space Science Institute (ISSI) in Bern, through ISSI International Team project \#545 ``Observe Local Think Global: What Solar Observations can Teach us about Multiphase Plasmas across Physical Scales''.

\bibliographystyle{apj}
\bibliography{bibfile.bib}

\end{document}